%% file: ms.tex
\documentclass[preprint2]{aastex}

\usepackage[]{natbib}
\usepackage{graphicx}
\usepackage{amssymb}
\newcommand{\etacar}{$\eta$~Car}
\newcommand{\ec}{\etacar}
\newcommand{\centconst}{CCE}

\newcommand{\myemail}{kenji@milkyway.gsfc.nasa.gov}

\input{macro_etacar.inp}

\slugcomment{2nd revision according to the referee's comment}

\shorttitle{X-ray Spectral Variation of \etacar}
\shortauthors{Hamaguchi et al.}

\begin{document}

\title{X-ray Spectral Variation of \etacar\  through the 2003 X-ray Minimum}

\author{Kenji Hamaguchi\altaffilmark{1,2}, Michael F. Corcoran\altaffilmark{1,2},
Theodore Gull\altaffilmark{3}, Kazunori Ishibashi\altaffilmark{4}, Julian M. Pittard\altaffilmark{5},
D. John Hillier\altaffilmark{6},
Augusto Damineli\altaffilmark{7}, Kris Davidson\altaffilmark{8}, Krister E. Nielsen\altaffilmark{3,9}, 
and Gladys Vieira Kober\altaffilmark{3,9}}

\altaffiltext{1}{CRESST and X-ray Astrophysics Laboratory NASA/GSFC ,
Greenbelt, MD 20771}

\altaffiltext{2}{Universities Space Research Association, 
10211 Wincopin Circle, Suite 500, Columbia, MD 21044}

\altaffiltext{3}{Astrophysics Science Division, NASA Goddard Space Flight Center,
Greenbelt, MD 20771}

\altaffiltext{4}{Kavli Institute for Astrophysics and Space Research, Massachusetts Institute of Technology, 77 Massachusetts Ave. NE80-6035, Cambridge, MA 02139}

\altaffiltext{5}{School of Physics \& Astronomy, The University of Leeds, Woodhouse Lane, Leeds, LS2 9JT, UK}

\altaffiltext{6}{Department of Physics and Astronomy, University of Pittsburgh, 3941 O'Hara Street, 
Pittsburgh, PA 15260}

\altaffiltext{7}{Departamento de Astronomia do IAGUSP,
R. do Matao 1226, 05508-900, S{\~ a}o Paulo, Brazil}

\altaffiltext{8}{Astronomy Department, University of Minnesota, 116 Church Street SE, Minneapolis,
MN 55455}

\altaffiltext{9}{Catholic University of America, Washington DC20064}

\altaffiltext{}{{\tt Mail to:} \myemail}

\begin{abstract}
We report the results of an X-ray observing campaign on the massive, 
evolved star \etacar,  concentrating on the 2003 X-ray minimum as seen by the \XMM\  observatory. These are the first spatially-resolved X-ray monitoring observations of the stellar X-ray spectrum during the minimum.
The hard X-ray emission, believed to be associated with the collision of \ec's wind with the wind from a massive companion star, varied strongly in flux on timescales of days, but not significantly on timescales of hours.  The lowest X-ray flux in the $2-10$ keV band seen by \XMM\  was only 0.7\% of the maximum seen by \RXTE\  just before the X-ray minimum.
In the latter half of the minimum, the flux increased by a factor of 5 from the lowest observed value,
indicating that
the X-ray minimum has two states.
The slope of the X-ray continuum above 5 keV did not vary in any observation, which suggests that the electron temperature of the hottest plasma  associated with the stellar source did not vary significantly  at any phase. 
Through the minimum, the absorption to the stellar source increased by a factor of $5-10$ to \NH\  $\sim$3--4$\times$10$^{23}$~\UNITNH.
The thermal Fe XXV emission line showed significant excesses on both the red and 
blue sides of the line outside the minimum and exhibited an extreme red excess during the minimum.
The Fe fluorescence line at 6.4~keV increased in equivalent width from 100~eV outside the minimum
to 200~eV during the minimum.
The small equivalent widths of the Fe fluorescence line
suggests small fluorescence yield in the companion's low-density wind.
The lack of variation in the plasma temperature is consistent with the eclipse of the X-ray plasma during the minimum, perhaps by a clumpy wind from the primary star, although
the deformation of the \ion{Fe}{25} profile and the relatively weak fluorescence Fe line intensity 
during the minimum may suggest an intrinsic fading of the X-ray emissivity.  
The drop in the colliding wind X-ray emission revealed the presence of an additional X-ray 
component which exhibited no variation on timescales of weeks to years.
This new component has relatively cool temperature (\KT\  $\sim$1~keV),
moderate \NH\  ($\sim$5$\times$10$^{22}$~\UNITNH), large intrinsic luminosity (\LX\  $\sim$10$^{34}$~\UNITLUMI) and a size $\lesssim$1\ARCSEC\  (2300~AU at 2.3~kpc).
This component may be produced by the collision of 
high speed outflows at $v \sim$1000$-$2000~\UNITVEL\  from \etacar\  with ambient gas within a few thousand AU from the star.
\end{abstract}

\keywords{Stars: individual (\etacar) --- stars: early-type --- stars: winds, outflows ---  
binaries: general --- X-rays: stars}

\section{Introduction}

\ec\  is a violently unstable, extremely luminous object and a key tracer of evolution of stars 
in the upper portion of the Hertzsprung-Russell diagram \citep{Davidson1997}.
The star is believed to have had
an initial mass of $\gtrsim$150~\UNITSOLARMASS\  \citep{Hillier2001}.
It is currently in a short, poorly understood evolutionary stage, 
known as the Luminous Blue Variable (LBV) phase, 
which is thought to occur near the onset of pulsational instabilities.
It provides a convenient laboratory to study how extremely luminous, massive stars evolve and 
how they shape their environments both geometrically and chemically.  

\ec\  is best known for an extraordinarily powerful eruption in 1843 which sent $\sim$12 M$_{\odot}$
\citep{Smith2003b}
of its atmosphere into space, creating a beautiful bipolar nebulosity called the Homunculus 
around the star.
\ec\  also had a minor eruption in 1890, which produced
a small bipolar nebula inside the Homunculus nebula
\citep[the ``Little Homunculus",][]{Ishibashi2003}. 
The star still exhibits a strong mass loss
\citep[10$^{-4}-$10$^{-3}$ \UNITSOLARMASSYEAR][]{Davidson1995,Cox1995,Ishibashi1999,Hillier2001,Pittard2002},
preferentially in the polar direction \citep{Smith2003,Boekel2003}. 
The ejecta and the stellar wind are rich in helium and nitrogen and depleted in oxygen and carbon
\citep{VernerE2005a,Davidson1984,Hillier2001},
consistent with CNO processing.
\citet{Davidson2005} and \citet {Martin2004} suggested from the change of 
H${\alpha}$ and H${\beta}$ line profiles and the
increase of optical brightness that \ec\  is changing rapidly at the present time.

The past decade has witnessed an important change in our understanding of the star.  This began with the recognition that the strength of some
narrow emission lines, notably He I 10830\AA, vary predictably 
with a period of 5.52 years \citep{Damineli1996}.
Along with observations in infrared \citep{Whitelock1994, Feast2001},
$mm$ \citep{Cox1995}, $cm$ \citep{Duncan1995}, and optical  \citep{Genderen1999} wavelengths, observations of the X-ray emission from the star have played a
key role.  \ROSAT\  observations in 1992 first showed a variation which
appeared to be correlated with Damineli's emission line variations
(Corcoran et al.  1995), and subsequently the $2-10$ keV lightcurve of
the star obtained by \RXTE\  \citep{Ishibashi1999} showed in  detail the X-ray
variation culminating in a swift, unstable rise to maximum and steep
fall to minimum lasting for $\sim$3 months,
i.e. the same time interval as Damineli's spectroscopic minimum.
These observations clearly show that the star varies in a fundamental way
every 5.54 years, and strongly suggest that \ec\  is two stars, not one.
In this binary model, most of the dramatic, pan-chromatic changes are now believed to 
be produced by the interaction of the UV flux and wind from a companion star with the wind of \etacar.
Variations in the X-ray region are produced by a wind-wind collision (WWC) of the wind from \etacar\  
with the wind of the companion star.
A current guess at the system parameters describes
a massive hot companion with 
$M_{c} \sim$30 \UNITSOLARMASS\  plus a brighter primary star with 
$M_{\eta} \geq$80 \UNITSOLARMASS\ 
in a highly eccentric ($e > 0.9$), $5.54$ year orbit
(\citealt{Corcoran2001b}, see also \citealt{Davidson1999}).

\input{tab1}

During the 1997/98 X-ray minimum, \ec\  was the target
of every available X-ray observatory, including \RXTE\  \citep{Ishibashi1999,
Corcoran2001b}, \ASCA\  \citep{Corcoran2000}, and
\SAX\  \citep{Viotti2002}.
The \ASCA\  satellite detected hard X-ray emission from \etacar\  \citep{Corcoran2000}, 
which was characterized by slightly smaller \NH\  than the pre-minimum state with reduced plasma 
emission measure (\EM).
The limited spatial resolution of \ASCA, however, left the possibility that 
the observed emission was due to contamination by unresolved nearby sources.
The \SAX\  satellite observed \etacar\  just after the recovery of the X-ray emission
\citep{Viotti2002}.
The spectrum showed strong absorption of 
\NH $\sim$1.5$\times$10$^{23}$~\UNITNH\  unlike
the \ASCA\  spectrum during the minimum, which had an \NH\ $\sim$3$\times$10$^{22}$~\UNITNH.
The overall X-ray brightness variations observed with \RXTE\
are explained well with the colliding wind mechanism
\citep{Pittard1998, Pittard2002}.
However, 
key properties of the X-ray lightcurve were still poorly understood: dramatic changes in X-ray
flux as the emission increases to maximum, the variation of the absorbing material in front of the X-ray emitting region near the X-ray minimum, the nature of the rapid fall from
X-ray maximum to X-ray minimum, the excess in \NH\ after recovery.

The minimum which occurred in mid-2003 was among the best observed astronomical events of all time. 
A key part of this campaign were the detailed X-ray observations obtained with \RXTE, \CHANDRA, and \XMM. \RXTE\ again provided crucial monitoring of the daily changes in X-ray flux, while \CHANDRA\  provided previously-unobtainable monitoring of the X-ray emission line dynamics at key phases of the binary period.  
\XMM\ obtained critical measurements of the spectrum of the source during the low-flux state when
observations with \RXTE\  or \CHANDRA\ are difficult.

This paper describes the overall change of the X-ray spectrum of \ec\  around the 2003 minimum
as measured by \XMM\  \citep{Jansen2001}, supplemented where necessary with key observations 
with \CHANDRA\  \citep{Weisskopf2002}.
We try to understand the X-ray emission and absorption mechanism and the cause of the 
X-ray minimum, from a comparison of derived spectral properties such as \KT, \NH\  and emission 
line strengths in these observations and earlier \ROSAT\  and \ASCA\  observations during the 
1992 and 1997/98 minima.

The paper is comprised of the following sections. 
Section~2 describes the observations and method of the data reduction,
and includes a description of problems of unresolved emission components around the 
central source in the extracted \XMM\  source events.
Section~3 describes imaging and timing analyses of the \XMM\  data.
Section~4 describes the analysis of the X-ray spectra during the 2003 minimum.
Section~5 compares the X-ray spectra during the 2003 minimum
to the previous minima in 1998 and 1992.
We discuss our results in section~6, and in 
section~7 we summarize our conclusions.

\begin{figure}[t]
\plotone{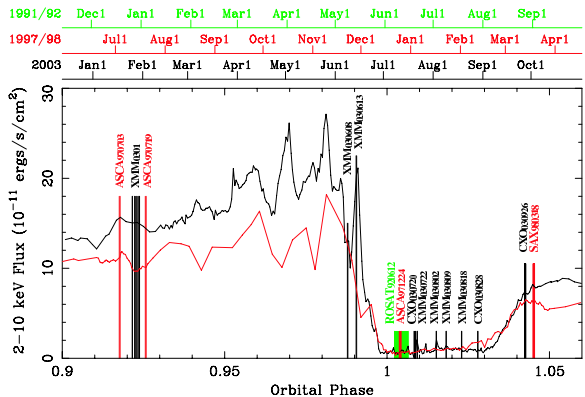}
\caption{\RXTE\  light curves (solid lines, {\it black}: 2003, {\it red}: 1997/98)
around the X-ray minimum 
with the timings of \XMM, \CHANDRA, \ASCA, \SAX, and \ROSAT\  pointed observations 
({\it black}: 2003, {\it red}: 1997/98, {\it green}: 1992).
Multiple \ROSAT\  observations between 1.00 to 1.01 were summed together for ROSAT$_{\rm 920612}$.
\label{fig:obstiming}
}
\end{figure}

\section{Observations \& Data Reduction}

In 2003, \XMM\  observed \etacar\ a total of 11 times,
5 times in January 2003 (about 6 months before the onset of the 2003 minimum),
twice in June 2003 (weeks before the onset of the 2003 minimum, near the X-ray
maximum) and
four times during the 2003 minimum in July and August 2003 prior to the recovery, 
which started near 2003 September 3. 
Figure~\ref{fig:obstiming} displays these observing times on the 
\RXTE\  lightcurve \citep{Corcoran2005} for context.
We also re-analyzed two \XMM\  observations from 2000,
whose results were previously published by \citet{Leutenegger2003}.
We thus used a total of 13 \XMM\  observations in our analysis,
as  listed in Table \ref{tbl:obslogs}.
Throughout this paper, individual \XMM\  observations are designated XMM, 
subscripted with the year, month and day of the observation.
To improve statistics, we combined data from two observations in 2000 and combined data from 
five observations in 2003 January because these observations were close together in time, 
and did not show any significant variabilty among them.

\XMM\  is composed of three nested Wolter I-type X-ray telescopes \citep{Aschenbach2000} with the European Photon Imaging Camera (EPIC) CCD detectors
(PN, MOS1 and MOS2) in their focal planes \citep[][]{Struder2001, Turner2001}.
In all the observations, \etacar\  was on-axis but the observations were obtained 
with different satellite roll angles.
The EPIC instrumental modes are listed in Table~\ref{tbl:obslogs}.
The \XMM\  spectra suffered significant photon pileup in the data taken in prime full window (PFW)
mode outside the 2003 minimum when the star was bright in the $2-10$ keV band. 
We did not use these data sets in our analysis.
\XMM\  is also equipped with the Reflection Grating Spectrometer \citep[RGS;][]{Herder2001},
but the RGS has limited sensitivity above $\sim$1~keV,
where emission from the WWC region dominates.
We therefore did not use the RGS data.

The analysis of the \XMM\  data was performed with version 5.4.1 of the 
SAS\footnote{http://xmm.vilspa.esa.es/external/xmm\_sw\_cal/sas\_frame.shtml}
software package and version 5.2 of the HEASoft\footnote{http://heasarc.gsfc.nasa.gov/docs/software/lheasoft/} analysis package.
The Observation Data Files (ODF) data were processed using the SAS scripts ``emchain" 
and ``epchain."
We removed events close to hot pixels or outside the field of view, and selected
events with pattern $\leqq$4 for EPIC PN spectral analysis and pattern $\leqq$12 for 
timing analysis of the EPIC PN data and spectral and timing analysis of the EPIC MOS data.
Most of these observations luckily avoided high background periods
and the instrumental background is negligible in any observations within the 0.3$-$10~keV band.
We did not need to reject any high background periods because the strong emission
from \etacar\  greatly surpassed the observed background levels.

\begin{figure*}[t]
\epsscale{2.15}
\plotone{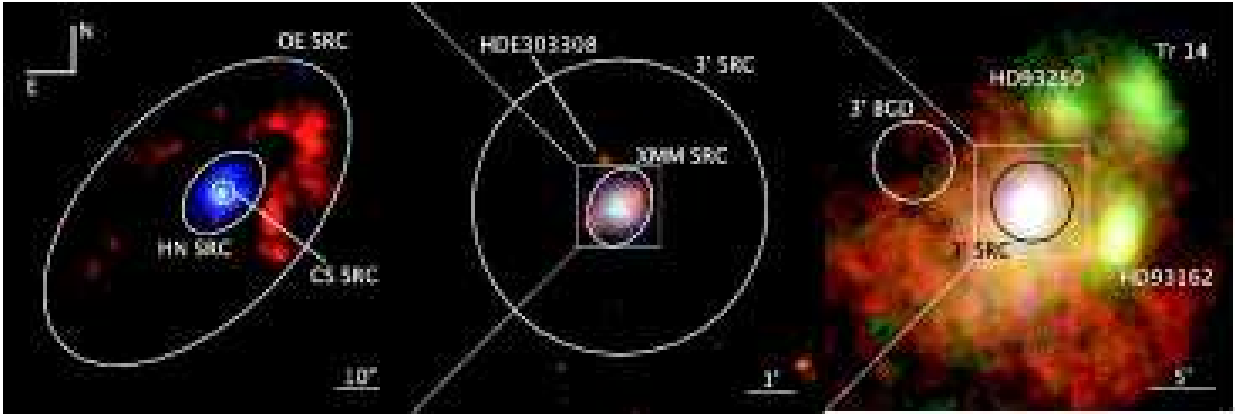}
\caption{True color images of \etacar.  {\it left}: \CHANDRA\  HETG 0th-order data from CXO$_{030720}$ 
(red: 0.2$-$1~keV, green: 1$-$3~keV, blue: 3$-$10~keV); 
{\it middle}: \XMM\  image accumulated from all the available MOS data taken in prime full window mode
(red: 0.2$-$1~keV, green: 1$-$3~keV, blue: 3$-$9~keV); {\it right}: 
\ASCA\  GIS data from ASCA$_{971224}$ 
(red: 0.8$-$1~keV, green: 1$-$3~keV, blue: 3$-$10~keV).
The color in the \ASCA\  image has been chosen to highlight the diffuse emission.
Solid lines show source (SRC) or background (BGD) regions 
(Designations: OE: Outer Ejecta, HN: Homunculus Nebula, CS: Central Source).
The \CHANDRA\  and \ASCA\  images were smoothed with a Gaussian with $\sigma$ = 1.5 pixel,
while the \XMM\  image was not smoothed.
\label{fig:obsimages}
}
\end{figure*}

Due to the limited spatial resolution of \XMM, the extracted \etacar\  data inevitably include
emission from a number of sources (see Figure~\ref{fig:obsimages}).
The extracted spectrum is usually dominated by hard X-rays 
from the wind-wind collision in \etacar, which varies from observation to observation
\citep{Corcoran2000,Viotti2002}.
The Homunculus Nebula emits weak diffuse reflected X-rays from the central source, 
which can be spatially resolved 
only with \CHANDRA\  and only during the minimum \citep{Corcoran2004}.
The outer ejecta extending about $1'$ beyond the Homunculus Nebula emits diffuse non-variable
X-rays below $\sim$1~keV with a ring- or shell-like morphology \citep{Seward2001,Weis2004}.
To correct the \XMM\  spectra for emission from the outer ejecta and the Homunculus Nebula,
we used three \CHANDRA\  data sets from the 2003 X-ray observing 
campaign, two observations
during the 2003 minimum and an observation just after the 2003 minimum.
These \CHANDRA\  observations spatially resolve the outer ejecta emission from the emission of the stellar 
source and so allow us to determine the spectrum of the outer emission, which we can then use 
to help model the \XMM\  spectra.

The log of \CHANDRA\  observations and the timing of the observations are also given in Table~\ref{tbl:obslogs} and Figure~\ref{fig:obstiming}.
\CHANDRA\  observations are designated CXO, subscripted with the year, month and day of the observation, similar to our designation of the \XMM\  observations.
The \CHANDRA\  observations were obtained using the Advanced CCD Imaging 
Spectrometer detector using the Spectrometer array (ACIS-S) either with or without the
high energy transmission grating (HETG).
For the grating data, 0th-order photon events were used for CXO$_{\rm 030720}$ 
(during the X-ray minimum, when the source was too faint at 1st order),
and 1st-order photon events were used for CXO$_{\rm 030926}$ after the recovery
since the central source had severe pile-up in the 0th-order data.
The observation CXO$_{\rm 030828}$, taken during the minimum with the ACIS-S imaging array
with no grating, also suffered mild pile-up ($\sim4\%$).
The analysis of the \CHANDRA\  data was performed with the 
CIAO\footnote{http://cxc.harvard.edu/ciao/} software package,
version 2.3 (CALDB version 2.22), version 3.0 (CALDB version 2.23) and version 3.0.2 (CALDB version 2.26) for sequences 200216, 200237, and 200217 respectively.
We also used version 5.2 of the HEAsoft package, and followed the recommendations of the analysis science 
thread\footnote{\url{http://cxc.harvard.edu/ciao/threads}}.  A full analysis of the Chandra spectra will be published separately (Corcoran et al. 2007, in preparation).

\section{X-ray Images and Time Variability}
\label{sec:xrayimglc}

\begin{figure*}
\plotone{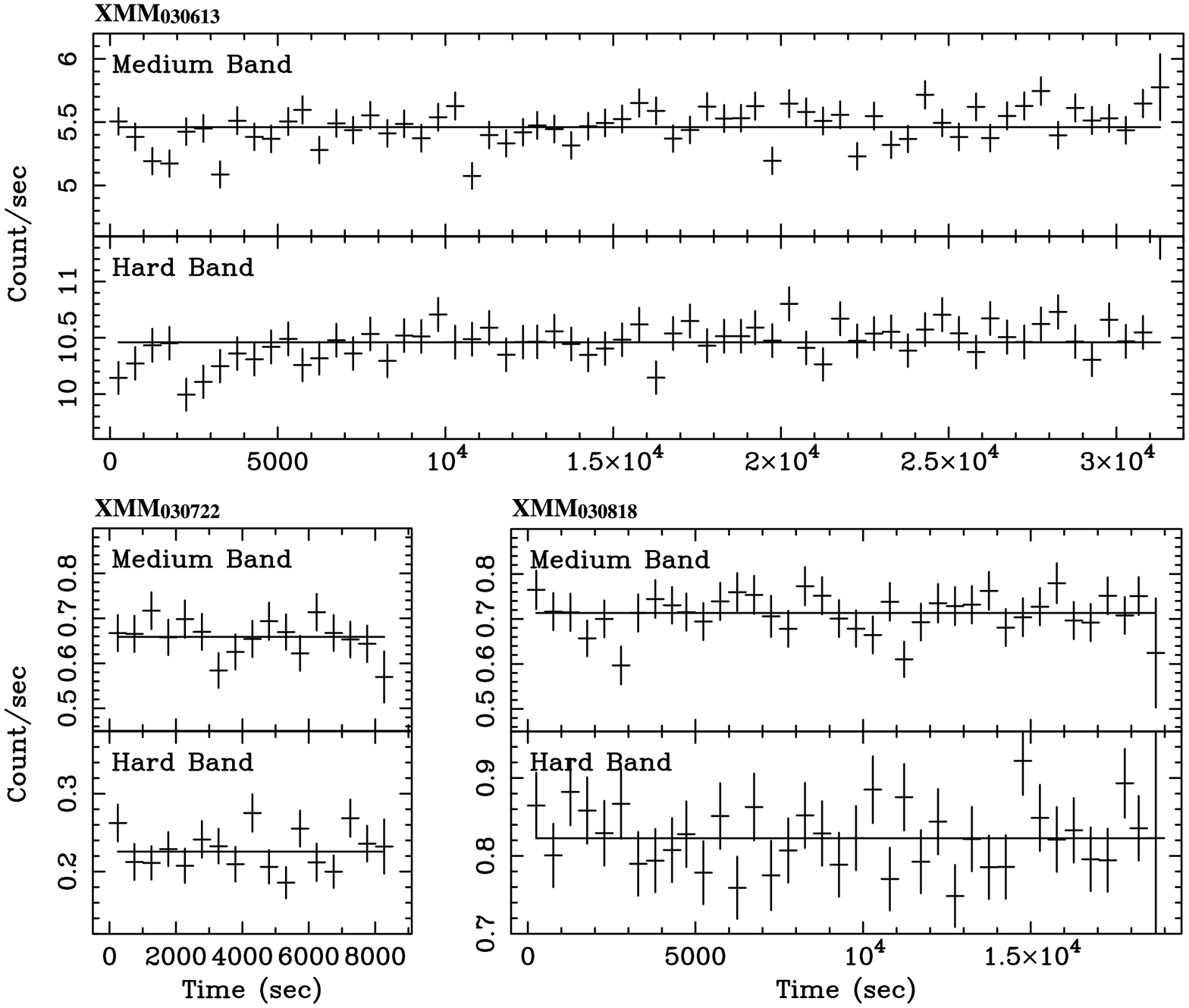}
\caption{Background subtracted EPIC PN~+~MOS light curves from XMM$_{\rm 030613}$ ({\it top}), XMM$_{\rm 030722}$ ({\it bottom left}),
and XMM$_{\rm 030818}$ ({\it bottom right}). Top and bottom graphs in each panel show the medium (1--4~keV)
and hard (4--10~keV) bands, respectively. A solid line in each panel shows the best-fit
constant model.
\label{fig:lcexp}
}
\end{figure*}

\input{tab2}

All \XMM\  images similarly showed emission from the central source and 
the outer ejecta (Figure \ref{fig:obsimages}).
Because these \XMM\  images cannot completely separate emission from the outer ejecta and central source
nor resolve the faint X-ray Homunculus Nebula \citep{Corcoran2004},
we used an ellipse of 50\ARCSEC\  $\times$ \mbox{37\FARCS5} as the source region  
(called ``XMM SRC" in Figure \ref{fig:obsimages}),
which includes all these emission components
but excludes nearby X-ray luminous stars (such as HDE~303308) more than $\sim1'$ distant.
For some small window mode MOS observations, the south-east tip of the ellipse fell outside the small window.
We determined background using regions from apparently source-free regions 
on the same CCD chip.

Table~\ref{tbl:cntslc} shows detected photon count rates in the soft (0.3--1~keV), 
medium (1--4~keV), and hard (4--10~keV) bands in each detector for each observation.
The soft band mainly includes emission from the outer ejecta, while
the medium band 
includes emission from an additional source which we call the central constant emission component (CCE, see \S\ref{subsec:nonvarcent}), 
and the medium band outside the minimum and the hard band includes the WWC emission
(see \S\ref{sec:varspec}).
In the soft band, the MOS2 count rates are 10\% lower in XMM$_{\rm 0007}$
than during the 2003 minimum because of the use of a smaller MOS2 
source region for the XMM$_{\rm 0007}$ data.
The soft-band PN count rates are slightly higher outside minimum because the spectral response of the PN allows photons from the bright hard source to contaminate the soft band.
For example,
we estimate the soft band contamination due to the hard photons in XMM$_{\rm 0007}$ to be 
$\sim$0.08~\UNITCPS,
which makes the corrected PN count rates in the soft band 1.06~\UNITCPS\ in this observation.
This means that the emission from the outer ejecta
was constant to $\lesssim$2\% to better than 90\% confidence over a half-year interval.
On the other hand, both the medium and hard band count rates decreased dramatically
to $\sim$1--5\% of the maximum brightness seen by \XMM.

We constructed light curves with 500~s time bins, combining all the available MOS and PN data.
In Figure~\ref{fig:lcexp}, we show three sample light curves, representing variations before the
minimum (XMM$_{\rm 030613}$) and during the minimum (XMM$_{\rm 030722}$, XMM$_{\rm 030818}$).
In general, these light curves 
showed no significant variation in most energy bands at $>$90\% confidence 
(Table~\ref{tbl:cntslc}, Figure~\ref{fig:lcexp}).
The exceptions are the medium band light curves before the 2003 minimum.
The medium band light curve from the XMM$_{\rm 030608}$ observation showed an ``excess'' of $\sim$0.2~\UNITCPS\  ($\sim$4~\%) near the middle of the observation, while the medium band light curve from the  XMM$_{\rm 030613}$ observation showed a slight linear increase by 
$\sim0.54^{+0.21}_{-0.22}$ 
\UNITCPS\  day$^{-1}$ (1-$\sigma$) or $\sim$10\% day$^{-1}$.
The linear model is still not accepted above the 90\% confidence level ($\chi^{2}$/d.o.f. = 1.59),
due to small fluctuation on timescales of $\lesssim$500~s.
We note that 
the hard band light curve in XMM$_{\rm 030613}$
is also better reproduced by a linearly-increasing model with a slope of
$\sim0.78^{+0.32}_{-0.26}$
\UNITCPS\  day$^{-1}$, or $\sim$7.5\% day$^{-1}$,
with $\chi^{2}$/d.o.f. = 0.98.
This increase
is roughly consistent with
the \RXTE\  light curve ($\sim$15\% day$^{-1}$, see Figure~\ref{fig:obstiming}).

\section{X-ray Spectra}
\label{sec:xrayspec}
\begin{figure*}
\plotone{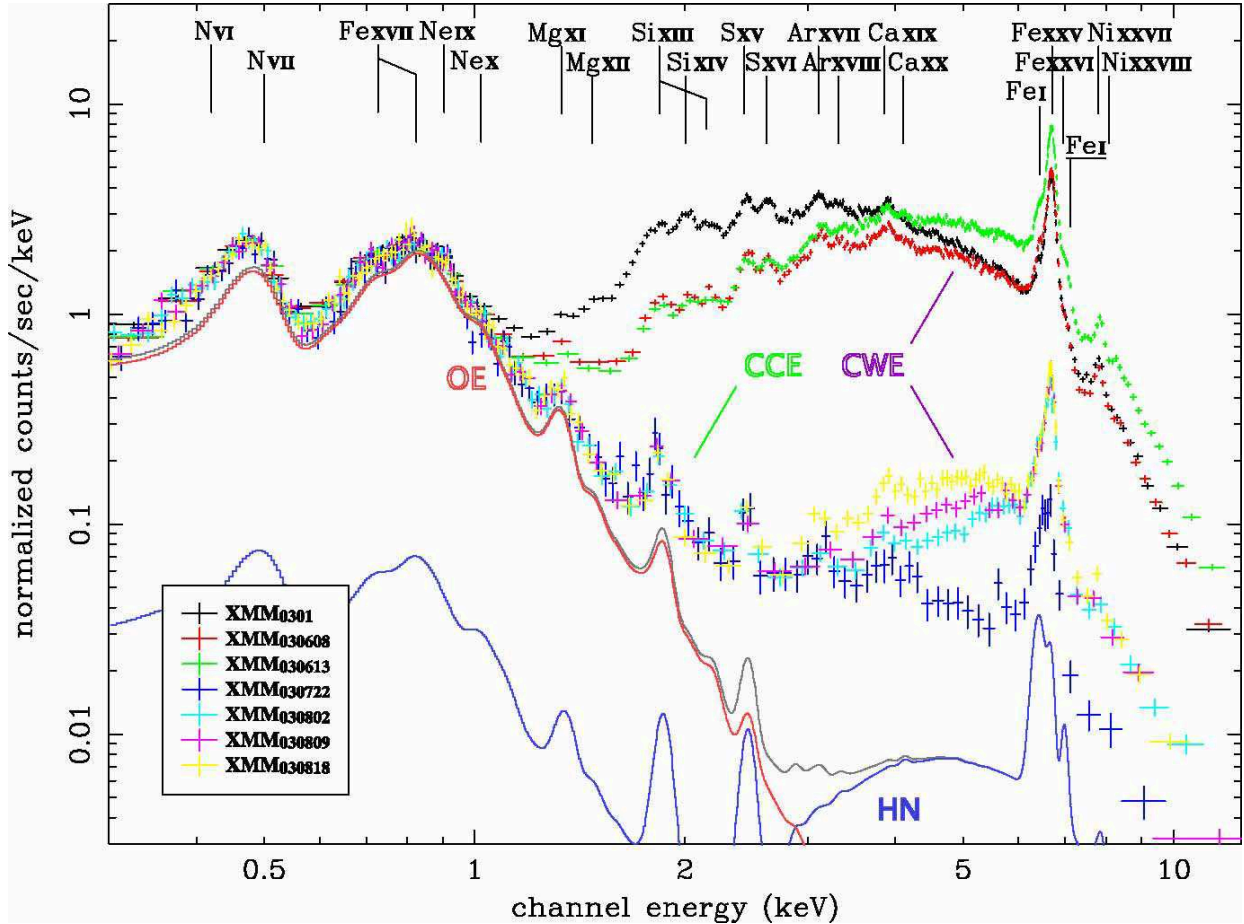}
\caption{\XMM\  EPIC PN spectra of \etacar\  in 2003.  The 
central source [central constant emission (\centconst) and colliding wind emission (CWE)] component,
the Homunculus Nebula (HN) and the outer ejecta components (OE) are marked.  
Identified emission lines are marked by lines at their rest energies.
\underline{\ion{Fe}{1}} marks the location of the Fe absorption edge.
The red, blue, and grey lines show \CHANDRA\  best-fit models of the outer ejecta
(see Figure~\ref{fig:cxoringspec}), the
Homunculus Nebula, and their sum, folded through the EPIC PN effective area.
\label{fig:specall}
}
\end{figure*}

For each \XMM\  observation, we produced EPIC PN and MOS spectra using 
the same source and background regions as in \S\ref{sec:xrayimglc}. 
Figure \ref{fig:specall} presents the EPIC PN spectra for all  2003 observations. The X-ray spectra are a combination of a variable hard WWC component,
along with non-variable emission from a number of components.
Consistent with Table~\ref{tbl:cntslc}, the spectra below 1~keV are almost constant through the
observations. 
Above 1~keV the spectra varied by a factor of 2$-$3 before the 2003 minimum,
then decreased dramatically by more than an order of magnitude after the onset of the 2003 minimum
in June, 
with small recovery by a factor of $\sim$3 at $E \gtrsim$4~keV during August.
Most spectra clearly showed lines of hydrogen- and helium- like ions of
Mg, Si, S, Ar, Ca, Fe, and Ni, and a fluorescence line from cold Fe.

In the following we examine first the non-variable emission, then use this emission to fully understand changes in the variable component.
In our analysis, we adopt abundances relative to solar abundances given by \citet{Anders1989}

\subsection{Non-Variable Components}

\subsubsection{Contribution of the Outer Ejecta and Homunculus Nebula to the X-ray Spectra}
\label{subsec:ringhomspec}
We use \CHANDRA\ observations to estimate the amount of contamination from the outer ejecta 
and the Homunculus Nebula
in the \XMM\ spectra.

We extracted the spectrum of the outer ejecta from the \CHANDRA\  CXO$_{\rm 030720}$ 
HETG 0th-order data, which has the smallest contamination from the central hard source.
We extracted the source spectrum from an 83\ARCSEC\  $\times$ 47\ARCSEC\  ellipse centered on the star (``OE SRC'' in  Figure~\ref{fig:obsimages}),
with its major axis parallel to the long axis of the extended outer X-ray nebula, but excluded
a 20\ARCSEC\  $\times$ 15\ARCSEC\  ellipse (``HN SRC'' in Figure~\ref{fig:obsimages}) 
centered on the central source and the Homunculus Nebula.  We extracted background spectra from
a nearby apparently source-free region. 
\etacar\ is embedded in soft diffuse emission from the Carina nebula, but its contribution to the \etacar\  spectrum is negligible.

\begin{figure}[h]
\epsscale{1.0}
\plotone{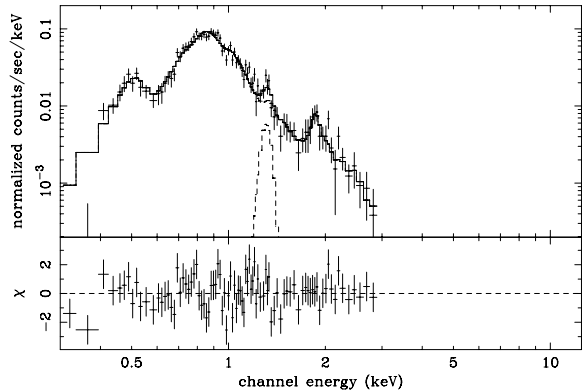}
\caption{\CHANDRA\  spectrum of the outer ejecta from CXO$_{\rm 030720}$ 
(OE SRC $-$ HN SRC in Figure~\ref{fig:obsimages}).
The solid line shows the best-fit absorbed 1T model and the dashed line depicts the 
Gaussian component for helium-like Mg.
\label{fig:cxoringspec}
}
\end{figure}

The spectrum from  the ``OE SRC'' region, excluding the spectrum from the ``HN SRC'' region, is shown in Figure~\ref{fig:cxoringspec}. This spectrum shows a strong emission line from nitrogen, 
which re-confirms the N enhancement in the outer ejecta found by a number of previous analyses,
(optical/UV: \citealt{Davidson1982}, X-rays: \citealt{Tsuboi1997}~[N/O $\sim$5],  \citealt{Leutenegger2003}~[N/O $>$9]).
This spectrum also shows a significant excess at 1.3 keV, the energy of the helium-like Mg line,
which seems to require a Mg overabundance of $\gtrsim$3~solar at \KT\  $\sim$0.6~keV
(though plasma with that temperature should emit a stronger hydrogen-like Mg line).
\citet{Leutenegger2003} also found a similarly strong Mg line in the \XMM\  grating spectrum, though they conjectured that some of the Mg emission might be contamination from the stellar X-ray source.  Our results suggest that the Mg line does in fact originate in the outer nebulosity.
We modeled the extracted, background-corrected source spectrum below 3~keV (where the emission 
from the extended source dominates) using 
a simple absorbed optically thin, thermal plasma model (WABS, \citealt{Morrison1983}; MEKAL, 
\citealt{Mewe1995}), including the
ACISABS\footnote{http://cxc.harvard.edu/cont-soft/software/ACISABS.1.1.html}
component to compensate for the
progressive low energy degradation of the quantum efficiency of the \CHANDRA\  ACIS detector.
We allowed the abundances to vary, but constrained all abundances to have the same value except for nitrogen, which was allowed to vary independently. We added a Gaussian with a fixed line centroid at 1.31~keV with no intrinsic broadening,
to account for the helium-like Mg line emission.
The model successfully reproduces the spectrum of the outer ejecta, yielding a reduced $\chi^{2}$ of 1.09 for 81 dof, 
with the best-fit parameters of
\KT\  $\sim$0.58~keV, \NH\ $\sim$7.2$\times$10$^{20}$~\UNITNH, with a VMEKAL normalization of 7.2$\times$10$^{-4}$,
elemental abundance of $\sim$85~solar for nitrogen, near solar abundance for the others,
and an excess flux of $\sim$1.5 $\times$10$^{-5}$~\UNITPFLUX\  for Mg.
The plasma temperature, $\sim0.6$ keV -- mainly constrained by continuum
above 1~keV -- agrees well with the highest plasma temperature derived from the
emission lines in the \XMM\  grating spectrum \citep{Leutenegger2003}.
However, the \NH\  is smaller than interstellar absorption to \etacar\  
($\gtrsim$3 $\times$ 10$^{21}$ \UNITNH, see \S2.2 of \citealt{Leutenegger2003}),
the nitrogen abundance is overestimated from earlier measurements,
and the best-fit model does not reproduce the Fe L and Ne lines near 1 keV
that are observed with the high resolution grating spectrum.
This is perhaps caused by assuming a simplistic 1T model for a spectrum which may have multiple cooler plasma components.
Therefore, this model reproduces CCD-resolution spectra 
($\gtrsim$100 eV at 1~keV) at energies above $\sim$0.4~keV but not higher-resolution grating spectra.

The reflected X-ray emission from the Homunculus Nebula did not vary significantly
between CXO$_{\rm 030720}$ and CXO$_{\rm 030828}$, an interval of 40~days \citep{Corcoran2004}.
Since this emission represents the accumulated X-rays reflected from the entire Homunculus Nebula,
any time variation of the central source 
shorter than the average light-travel time from the star to the reflecting site inside the Homunculus Nebula ($\sim$88~days) will be smeared out.
We assume that the emission from the Homunculus Nebula
did not vary between CXO$_{\rm 030720}$ and CXO$_{\rm 030828}$, which includes all
\XMM\  observations during the 2003 minimum (XMM$_{\rm 030722}$ -- XMM$_{\rm 030818}$).
To show the relative magnitude of the contamination from this reflected component, in 
Figure~\ref{fig:specall} we include a model fit of the \CHANDRA\  spectrum
of the reflected X-rays from the Homunculus Nebula,
which we extracted from a 20\ARCSEC\  $\times$ 15\ARCSEC\  ellipse centered on the star
(HN SRC in Figure~\ref{fig:obsimages}), excluding a 2\FARCS5 radius circle (CS SRC).

Though the X-ray intensity of the Homunculus in observations outside the 2003 minimum is less certain,
at these times the X-ray emission from the central source was so bright that
the X-ray contamination from the Homunculus is $\lesssim$1\% even if the reflected emission from the Homunculus brightened by a factor of $\sim$3.

\subsubsection{Discovery of a ``Constant'' Component Near the Central Source}
\label{subsec:nonvarcent}

\begin{figure}
\epsscale{1.0}
\plotone{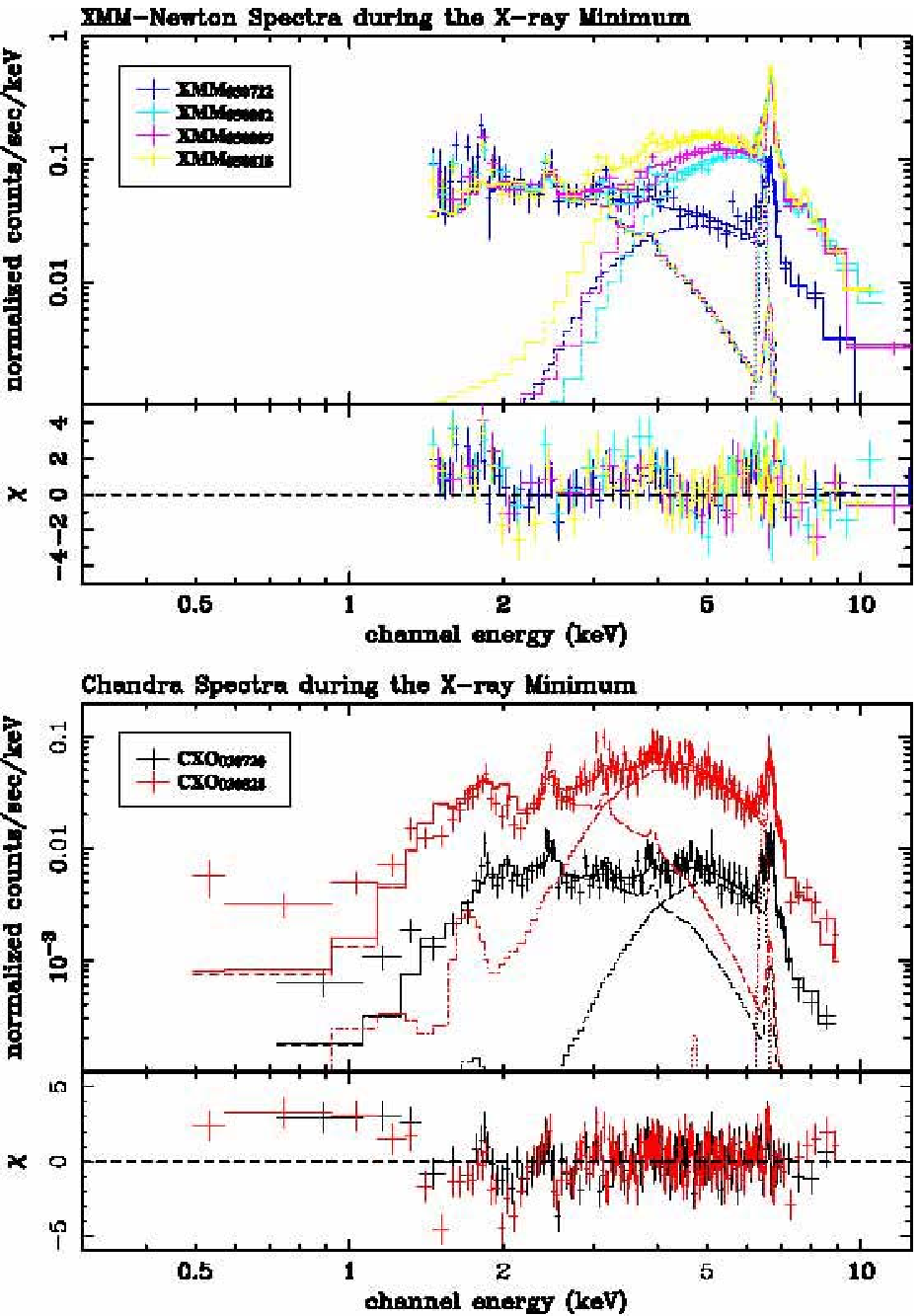}
\caption{{\it Top}: \XMM\  EPIC PN spectra during the 2003 minimum (XMM$_{\rm 030722}-$XMM$_{\rm 030818}$)
after subtracting emission
from the outer ejecta and Homunculus Nebula.  {\it Bottom}: \CHANDRA\  spectra of the central region in
CXO$_{\rm 030720}$ (black, HETG 0th-order) and CXO$_{\rm 030828}$ (red, ACIS-S).
The solid lines show the best-fit model of a simultaneous fitting of \XMM\  (PN and MOS)
and \CHANDRA\  spectra by an individually absorbed 2T model.
Barred, dot-bar, and dotted lines show the \centconst\  component, variable component, and
a Gaussian for the Fe K fluorescence line, respectively.
\label{fig:ringhom}
}
\end{figure}

\input{tab3}

Figure~\ref{fig:specall} 
depicts the \XMM\  EPIC PN spectra recorded through 2003, along with the best-fit spectral models of the outer ejecta, the X-ray Homunculus Nebula, 
and their sum.
Except for the excess below 1~keV,
which is produced by the poor absolute flux calibration of the EPIC 
PN\footnote{see the \XMM\  calibration report CAL-TN-0018-2-3},
the excess over the summed spectrum represents emission 
in a region (CS SRC in Figure~\ref{fig:obsimages}) within $2$\FARCS5 of the hard X-ray peak.
Apparent in Figure~\ref{fig:specall} is an excess over the model in the energy range $1-3$ keV.
This emission apparently did not vary in time with the harder emission, and had the same flux in the \CHANDRA\  spectra of the central region
in observations CXO$_{\rm 030720}$, CXO$_{\rm 030828}$, and CXO$_{\rm 030926}$
(see also Figure~\ref{fig:ringhom}), so that
the emission was securely constant for more than two months between July 20 -- September 26 2003.
We call this component the Central Constant Emission component, or ``\centconst''.

The \centconst\  shows lines of helium-like S and Si ions but no lines of hydrogen-like S and Si 
ions.
This indicates that the \centconst\  comes from relatively cool plasma.
To estimate the emission parameters of the \centconst,
we subtracted contributions of the outer ejecta and Homunculus Nebula from both the PN and MOS spectra
(see the top panel of Figure~\ref{fig:ringhom} which shows the PN spectra).
We ignored data bins below 1.4~keV, where the MOS and PN spectra have significant discrepancies
due to problems with the relative flux calibrations between the two instruments.
To better estimate the spectrum at low energies,
we also fit the \CHANDRA\  spectra of the central region from 
CXO$_{\rm 030720}$ and CXO$_{\rm 030828}$
(see the bottom panel of Figure~\ref{fig:ringhom}).
Because of the finer spatial resolution of \CHANDRA, these spectra are mostly free from contamination by emission from the outer ejecta.
Each \XMM\  and \CHANDRA\  spectrum was fit by a combination of a low and high temperature, optically thin thermal plasma model
with independent absorbing columns 
for each component, including a Gaussian line to account for the fluorescence iron line at 6.4~keV. The low temperature component accounts for the emission from the \centconst.

The parameters of the low temperature component 
were tied among all the spectra, and we assumed fixed solar abundances for all elements except for silicon and sulfur, whose abundances were allowed to vary simultaneously for all the spectra.
We allowed the parameters of the hard component (which represents the variable emission of the stellar source)  to vary among all the spectra, but we kept the abundances at the same (non-solar) values in each spectrum.
Table~\ref{tbl:centconst} shows the best-fit parameters for the constant component.
The spectrum can be fit with \KT\  $\sim$1.1~keV, \NH\  $\sim$5.0$\times$10$^{22}$ \UNITNH, 
with an absorption corrected luminosity of log \LX\  $\sim$34~\UNITLUMI, and 
low abundances for Si ($\sim$0.21~solar) and S ($\sim$0.47~solar).

The model, however, was not accepted at above 90\% confidence for the following reasons:
i) The \CHANDRA\  spectra show significant excess below $\sim$1~keV, perhaps
originating in foreground emission from the outer ejecta ``bridge'' \citep[][ see also Figure~\ref{fig:obsimages}]{Weis2004}.
The excess can be reproduced by including some flux from the outer ejecta 
using the model derived in \S\ref{subsec:ringhomspec}.  This slightly improves the reduced $\chi^{2}$ from 1.51 for 1205 to 1.50 for 1203 degrees of freedom.
The included flux is $\sim$0.4\% of the whole outer ejecta emission.
ii) The model overestimates the spectra near 2 and 2.7~keV,
where lines of hydrogen-like Si and S ions are,
and the model underestimates the spectra near 1.8 and 2.5~keV, where lines of helium-like Si and S ions are.
This suggests that the constant component includes plasma with temperature cooler than \KT\  $\sim$1~keV,
though an inconsistency near 2~keV between \XMM\  and \CHANDRA\  spectra could also be
caused by calibration uncertainties of the \CHANDRA\  effective area around the iridium M-edge.
Since cooler plasmas have lower emissivities in those lines,
our simple 1T model fitting for the \centconst\  component can underestimate the
elemental Si and S abundances.
iii) The model underestimates the \XMM\  spectra near 3$-$4~keV, possibly because the variable component
is a multiple temperature plasma (see \S\ref{sec:varspecallfit}).
iv) The Fe K line profile is complex and cannot be fit by a simple 1-temperature equilibrium model
(see \S\ref{sec:ironkprofile}). Nevertheless, the parameters 
given in Table \ref{tbl:centconst} should approximate the physical properties of the \centconst, except perhaps for the elemental abundances, which are more dependent on details of the model.

As seen in Figure \ref{fig:ringhom}, the emission between 1$-$3~keV in CXO$_{\rm 030720}$ comes predominantly from the \centconst\  component.
The 1$-$3~keV image in CXO$_{\rm 030720}$ was apparently point-like,
which restricts the projected plasma size to within $\sim$1\ARCSEC\  of \etacar,
the width of the \CHANDRA\  point-spread function.
This is equivalent to a projected distance $\lesssim$2300~AU at $d \sim$2.3~kpc,
suggesting that the hot gas which produces the \centconst\  component is inside the Little Homunculus Nebula \citep[which has a projected extent of $\sim\pm2''$, ][]{Ishibashi2003}.

\begin{figure*}
\epsscale{1.5}
\plotone{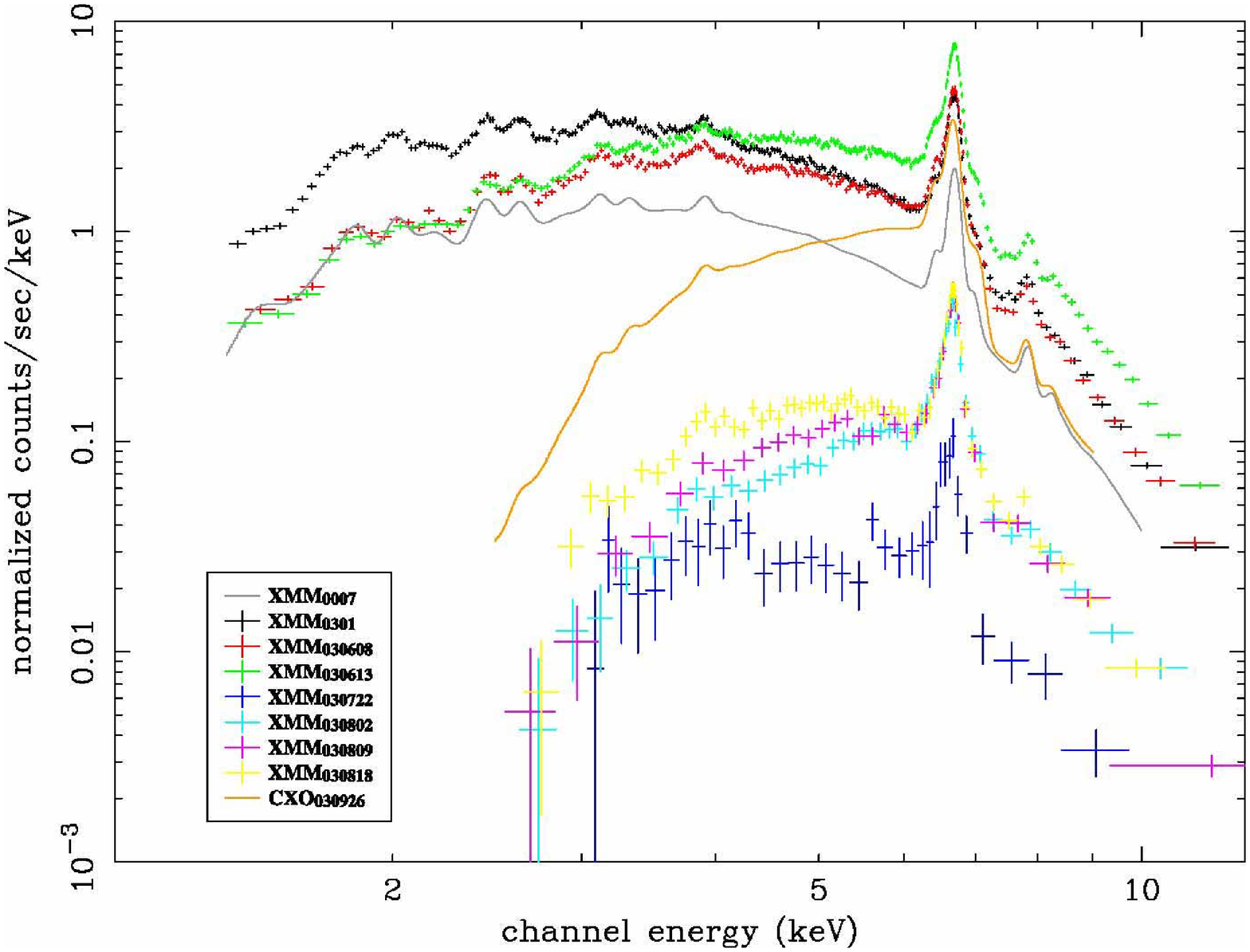}
\plotone{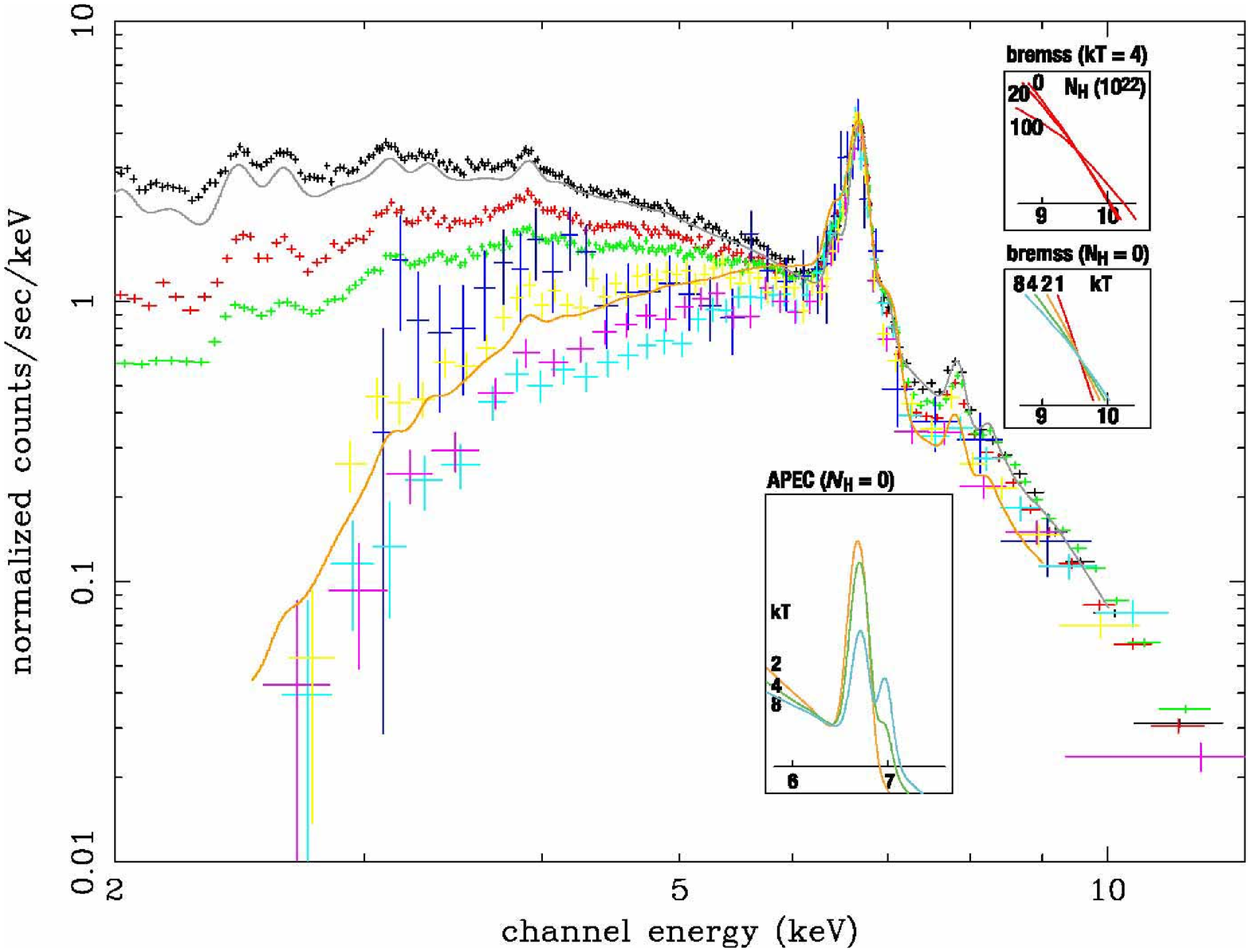}
\caption{{\it Top}: \XMM\  spectra of the variable emission.  {\it Bottom}:
\XMM\  spectra renormalized to the line intensity of the helium-like Fe line.
Best-fit models of the variable emission in XMM$_{\rm 0007}$ and CXO$_{\rm 030926}$,
convolved with the EPIC PN effective area, are also displayed in grey and orange.
Each color represents the same observation as in Figure \ref{fig:specall}.
Boxes in the bottom panel show the high energy slopes of the bremsstrahlung emission
at \KT\  =4~keV with \NH\  = 0, 2, and 10 $\times$10$^{23}$\UNITNH\ 
({\it top}), the bremsstrahlung emission at \KT\  =1, 2, 4, and 8~keV without absorption
({\it middle}),
and Fe line profile of the APEC model at \KT\  =2, 4, and 8~keV with abundance of 0.5~solar
({\it bottom}).
\label{fig:specnorm}
}
\end{figure*}

\subsection{The Variable Component}
\label{sec:varspec}

In order to determine the spectrum of the variable stellar component, we then subtracted from all the original spectra the recognized non-variable emission components 
-- the outer ejecta, the X-ray Homunculus Nebula, and the \centconst\  component.
The remaining emission represents the variable component associated 
with the stellar source, and is shown in Figure~\ref{fig:specnorm}.  The top panel shows the variable component ``corrected'' for the constant components we identified. For comparison, 
we overlaid on the top panel of Figure~\ref{fig:specnorm} the best-fit models for the XMM$_{\rm 0007}$ and CXO$_{\rm 030926}$ datasets.

\subsubsection{Variation of the Overall Spectral Shape}
\label{sec:varspecgenshape}

The bottom panel of Figure~\ref{fig:specnorm} shows the ``corrected'' variable components normalized at the \ion{Fe}{25} line energy 
to highlight changes in the shape of the spectra. 
From XMM$_{\rm 0007}$ to XMM$_{\rm 0301}$ (an interval of 2.5 years)
the X-ray flux increased by a factor of 2--3 without remarkable changes
in spectral shape.
Four months later (observation XMM$_{\rm 030608}$), the X-ray emission decreased only below 5~keV.
Five days later (observation XMM$_{\rm 030613}$), when \etacar\  was close to the peak of the last flare before 
the 2003 minimum \citep{Corcoran2005},
the hard X-rays above $\sim$3~keV increased by a factor of two.
The X-ray emission dropped by almost two orders of magnitude by
XMM$_{\rm 030722}$, the first observation after the 2003 minimum.
In XMM$_{\rm 030802}$, the hard band flux recovered by a factor of 3 without any significant change in the  soft band.
The hard band flux stayed the same in XMM$_{\rm 030802}$, XMM$_{\rm 030809}$, XMM$_{\rm 030818}$, and CXO$_{\rm 030828}$ (not shown), while the  soft band emission
slightly increased.
The \RXTE\ light curves \citep{Corcoran2005} also show a clear transition in the flux level and hardness
ratio between XMM$_{\rm 030722}$ and XMM$_{\rm 030802}$.
This means that 
the X-ray minimum has two states.
By CXO$_{\rm 030926}$, only the hard band flux recovered to the pre-minimum level.

The bottom panel in Figure~\ref{fig:specnorm} clearly shows that, throughout the observations,
the hard band slope above 7~keV did not change significantly, nor did
the ratio of hydrogen-like to helium-like Fe ion lines.
This means that the electron temperature of the hottest plasma did not change during the 2003 minimum.
On the other hand, the  soft band flux relative to the hard band decreased gradually
from the X-ray maximum through the minimum to the recovery.
This looks like an increase of the absorption column to the X-ray plasma though
the situation is somewhat more complicated, as the following sections describe. 
The 2003 minimum can be better described as an apparent decrease of the emission measure (\EM)
as suggested by earlier \ASCA\  observations \citep{Corcoran2000}, which means that either the amount of X-ray  emitting material has declined, or that the amount that is visible to the observer is smaller.

\input{tab4}

\subsubsection{1-Temperature Fit of the Entire Spectrum}
\label{sec:varspecallfit}

To quantitatively describe the spectral shapes and compare them with earlier results
obtained by \ASCA\  and \SAX\  (see \S\ref{sec:comppremin}), 
we fit each spectrum by an absorbed 1T thermal APEC\footnote{http://cxc.harvard.edu/atomdb/} 
model (which provided a better fit near the Fe K lines than the MEKAL model), 
with a Gaussian at 6.4~keV to account for the Fe~K fluorescence line.
The results are listed in Table~\ref{tbl:specallfit}.

Most of the fits were not acceptable at the 90\% confidence level.
One reason for the poor fit by the absorbed 1T models is that 
the strong Fe K$\alpha$ profile is quite complicated, as described in \S\ref{sec:ironkprofile}.
Another reason, in particular for the spectra before the 2003 minimum, is that
the plasma is really composed of multiple components with different temperatures as shown by
\citet{Corcoran2001a}, as their best-fit plasma temperatures \KT\ $\sim$4--5~keV 
do not account for the strong lines of helium-like S and Si ions seen in the \XMM\  spectra 
(which should be emitted from plasma with  \KT\ $\sim 1$~keV).
The plasma temperatures outside the 2003 minimum (\KT\ $\sim$4--5~keV) are consistent with 
earlier results \citep{Tsuboi1997,Corcoran2000,Viotti2002,Leutenegger2003},
but they are probably overestimated by our 1T model fit (see the next section).
The derived absorption column, which does not depend strongly on temperature for 
\KT\ $\sim$3$-$5~keV, gradually increased from 
\NH\ $\approx 5\times10^{22}$~\UNITNH\  to \NH\ $\approx 4\times10^{23}$~\UNITNH.

\subsubsection{Fits to the $E >$5 keV Spectrum}
\label{sec:ironkprofile}

\begin{figure*}
\epsscale{2.0}
\plotone{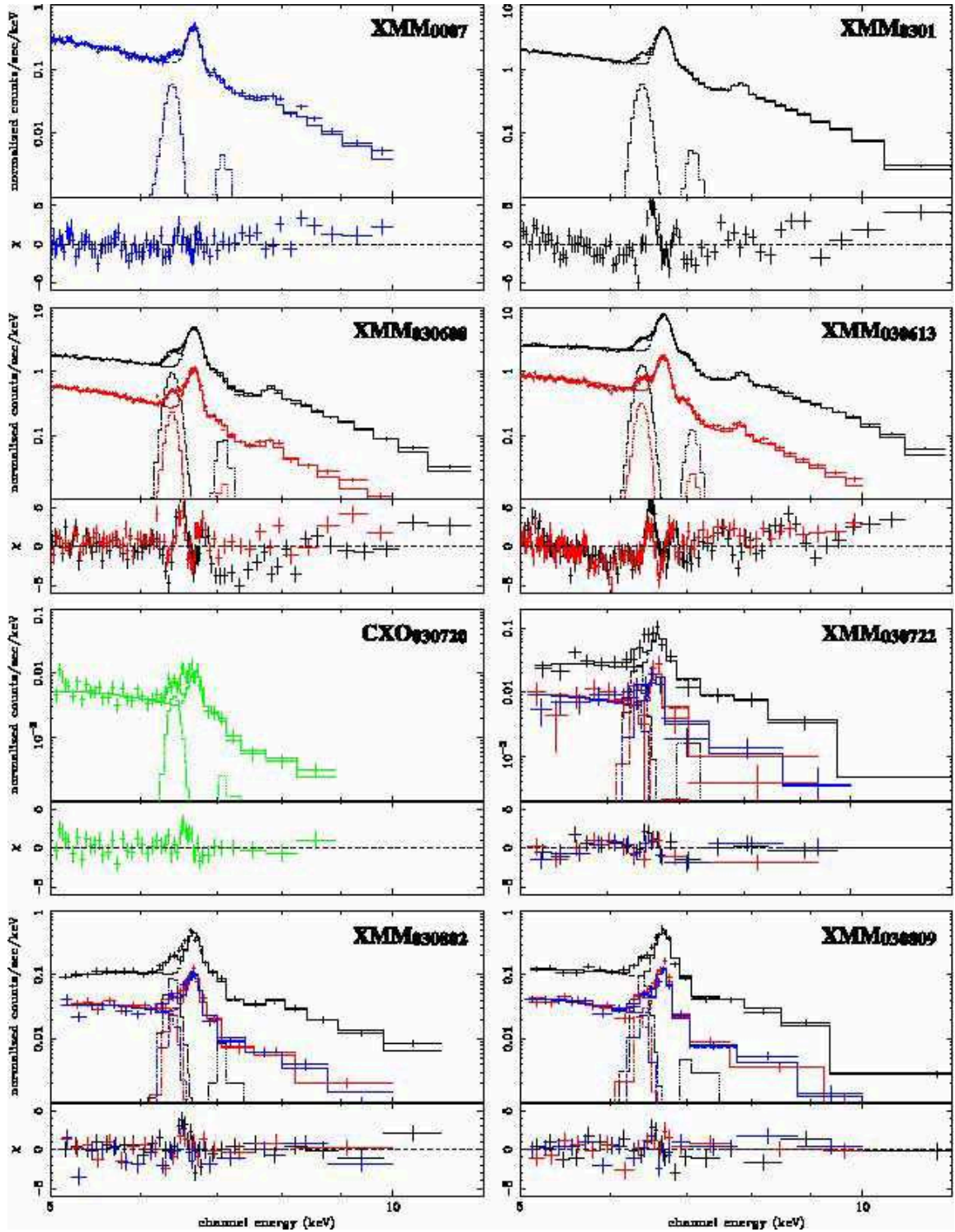}
\caption{Spectra above 5~keV: EPIC PN (black), EPIC MOS1 (red) and MOS2 (blue), ACIS (green)
and HETG (grey: +1st order, brown: $-$1st order).
The solid lines show the best-fit absorbed optically thin-thermal plasma model (APEC code) with 
Gaussian components for Fe K $\alpha$ and $\beta$.
\label{fig:xmmpnhard1}
}
\end{figure*}

\begin{figure*}
\plotone{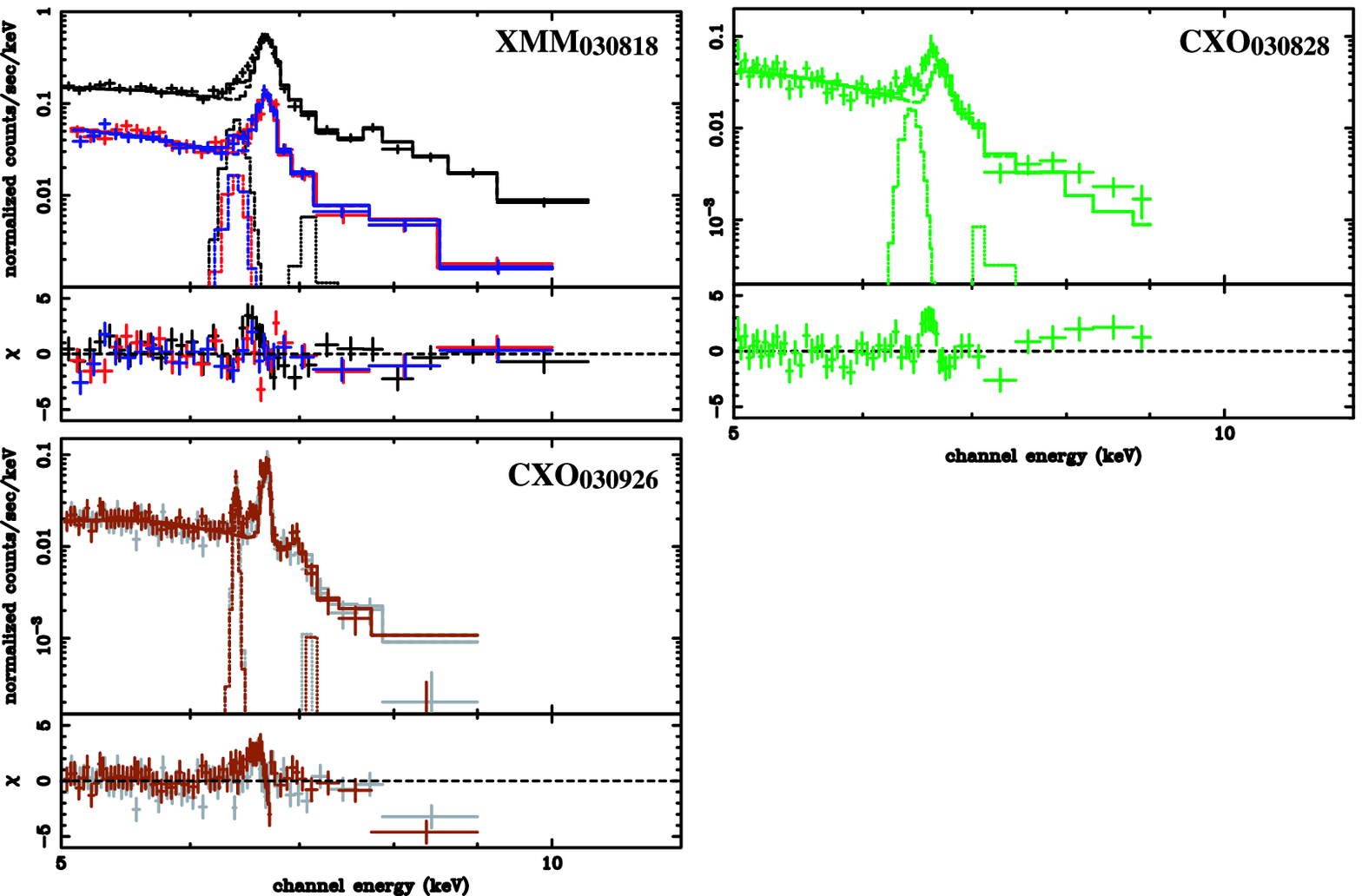}
\centerline{Fig. 8. --- Continued.}
\end{figure*}

The EPIC spectra marginally resolved emission lines below $\sim$5~keV
and could not reliably measure the neighboring continuum level.
Spectral modelling of this energy band with unconstrained elemental abundances, 
therefore, has intrinsic uncertainty.
On the other hand, the spectrum above 5~keV 
is relatively simple and provides an unambiguous measure of the hottest plasma 
from the continuum slope and Fe K line intensities.
To characterize the hottest plasma, we thus fit the spectra above 5~keV by a 
simple absorbed 1T model.

In each observation, we simultaneously fit all the available spectra, 
using an APEC thermal equilibrium model for the thermal emission.
We simultaneously fit the spectra of two observations, 
CXO$_{\rm 030720}$ and XMM$_{\rm 030722}$,
which were obtained during the 2003 minimum within 2 days of each other.
A preliminary fit to the bright phase spectra showed that most spectra could be fit with \KT\ =3.3 keV, a nickel abundance near 0.8 solar, with solar abundances for other elements.
Therefore, we fixed the nickel abundance at 0.8~solar and the temperature at 3.3~keV, and
we varied the Fe abundance, column density and emission measure.
We included 2 narrow Gaussian lines to account for lines of Fe K$\alpha$ and K$\beta$ fluorescence, fixing their centroids at 6.4~keV and 7.1~keV, with their line widths ($\sigma$) fixed at 0.01~keV, 
and the K$\beta$ line flux fixed at 11.3\% of the K$\alpha$ line flux.
We used photoelectric absorption cross-sections calculated by \citet{Balucinska1992}.
Table \ref{tbl:spechardfit} gives the best-fit parameters, and the spectra, along with their best-fit models, 
are shown in Figure \ref{fig:xmmpnhard1}.
The spectra except for XMM$_{\rm 0007}$ and CXOXMM$_{\rm 0307}$
accept the assumed model at $<$90\% confidence,
mostly because these spectra show several spectral features which are not fit by the model.

\input{tab5}

The most prominent of these features are excesses on the blue and red sides of the K-shell lines of helium-like iron.
Those excesses are not caused by poor energy or gain calibration, since there is good consistency in profile between 
the XMM$_{\rm 030613}$ spectrum and the near-contemporaneous \CHANDRA\  grating spectrum.
The blue excess could be explained if the K lines of \ion{Fe}{25} are Doppler broadened, with ${\Delta}v\sim$4000~\UNITVEL.  This is similar to the derived wind velocity of the companion star, $v\approx 3000~$\UNITVEL\  \citep{Pittard2002}.
The red excess, especially during the 2003 minimum
and CXOXMM$_{0307}$ and XMM$_{030802}$ when the X-ray flux was the weakest,
would require a
\ion{Fe}{25} Doppler shift of $\Delta v\sim$7000~\UNITVEL, and such
high velocities are not expected in  \etacar.
However, the red excess may be produced by unresolved emission lines of Fe in ionization stages below \ion{Fe}{25} in the 6.5--6.6 keV band. 
Interestingly, the \ASCA\  minimum spectrum in ASCA$_{971224}$ showed a similar excess
(see \S\ref{sec:minimumasca}),
while a \CHANDRA\  grating spectrum near apastron ($\phi \sim$0.53) 
showed a similar broad-band excess \citep{Pittard2002}.
Presence of lower-ionization iron lines would indicate that the hottest plasma is no longer in collisional ionization equilibrium at 3.3~keV, and that the electron temperature is higher than the ion temperature. 
Non-equilibrium ionization (NEI) effects have been claimed in the X-ray spectrum of WR 140, another colliding
wind system \citep{Pollock2005}.

Similarly, a hump-like spectral feature is perhaps present between 5$-$6.4~keV and becomes more noticeable during the early phase of the minima.
There are no specific  emission lines in this energy band, except a weak \ion{Ca}{19} line at 4.56 keV.
A similar feature is sometimes seen in the X-ray spectra of some AGNs \citep{Tanaka1995},
produced by gravitationally red-shifted material very near the central blackhole.
However, there is little evidence that 
\etacar\  houses a black hole because of
the lack of short term X-ray variability and relatively low X-ray luminosity.

A second feature is a hard ``tail'' at $\gtrsim$9~keV, best seen in the residuals in 
XMM$_{\rm 0007}$, XMM$_{\rm 0301}$, XMM$_{\rm 030608}$, and XMM$_{\rm 030613}$ 
(see Figure \ref{fig:xmmpnhard1}).
This feature is probably not an instrumental or background artifact: none of the spectra (except for 
CXO$_{\rm 030828}$) suffer photon pile-up, and source count rates are 
much higher than the background.
During the 2003 minimum the tail seems weaker or non-existent, though this may simply be an 
artifact since the source is weak at these times.
The slope above 9~keV in XMM$_{\rm 030613}$ can be fit by a bremsstrahlung model with \KT\  $\sim$10~keV 
or power-law model with $\Gamma\sim$ 2.3.
\citet{Viotti2004} measured a similar photon index for a hard tail extending up to 150~keV,
seen in a \SAX\  PDS spectrum in June 2000 ($\phi \sim$0.46),
but the flux between 13$-$20~keV, 1.4$\times$10$^{-11}$~\UNITFLUX,
is about three times larger than the extrapolation of the hard tail
we measured in XMM$_{\rm 030613}$.
The hard excess we see in the \XMM\  data may be from an extremely hot plasma 
in the colliding wind region, or from
non-thermal emission due to 1st-order Fermi acceleration at the wind contact surface, 
which would produce a population of relativistic electrons that can upscatter UV photons from 
the stellar photosphere and/or X-ray photons from the wind-wind shock.

We also clearly see an excess at $\sim$8.5~keV in the PN
spectra in observations XMM$_{\rm 0301}$ and XMM$_{\rm 030613}$.  This feature might be due to
emission from K shell lines of heavy metals like Cu and Zn.
Though the EPIC instrumental background
shows emission from those elements \citep{Struder2001}, inspection of the EPIC background data provided by the XMM project showed that
the instrumental background at this energy was negligible in spectra obtained outside the minimum.
This feature could indicate an overabundance of Cu or Zn, or perhaps be a part of the Cu edge at $\sim$9~keV
(which would also require a large Cu overabundance as well).  On the other hand, the feature might be produced  by absorption edges from \ion{Fe}{25} ($\sim$8.8~keV) and \ion{Fe}{26} at $\sim$9.3~keV
\citep{Lotz1967}.

Finally, the fourth feature is a marginal dip around 7~keV, which is seen in all the PN spectra,
but which is not so prominent in the MOS spectra.
This dip could be caused by an underestimate
of the line intensity of hydrogen-like Fe ions at 6.9~keV (perhaps due to an underestimate of the maximum plasma temperature)
or an underestimate of the strength of the Fe edge structure at 7.1~keV, 
perhaps indicating larger Fe absorption.
This dip is especially noticeable in the observation CXO$_{\rm 030926}$, and will be discussed more fully in the analysis of our \CHANDRA\  HETGS observations
(Corcoran et al. in preparation).

\subsubsection{Adding a Soft Component to Reproduce the Entire Spectrum}
\label{sec:softbandfit}

\begin{figure*}
\plotone{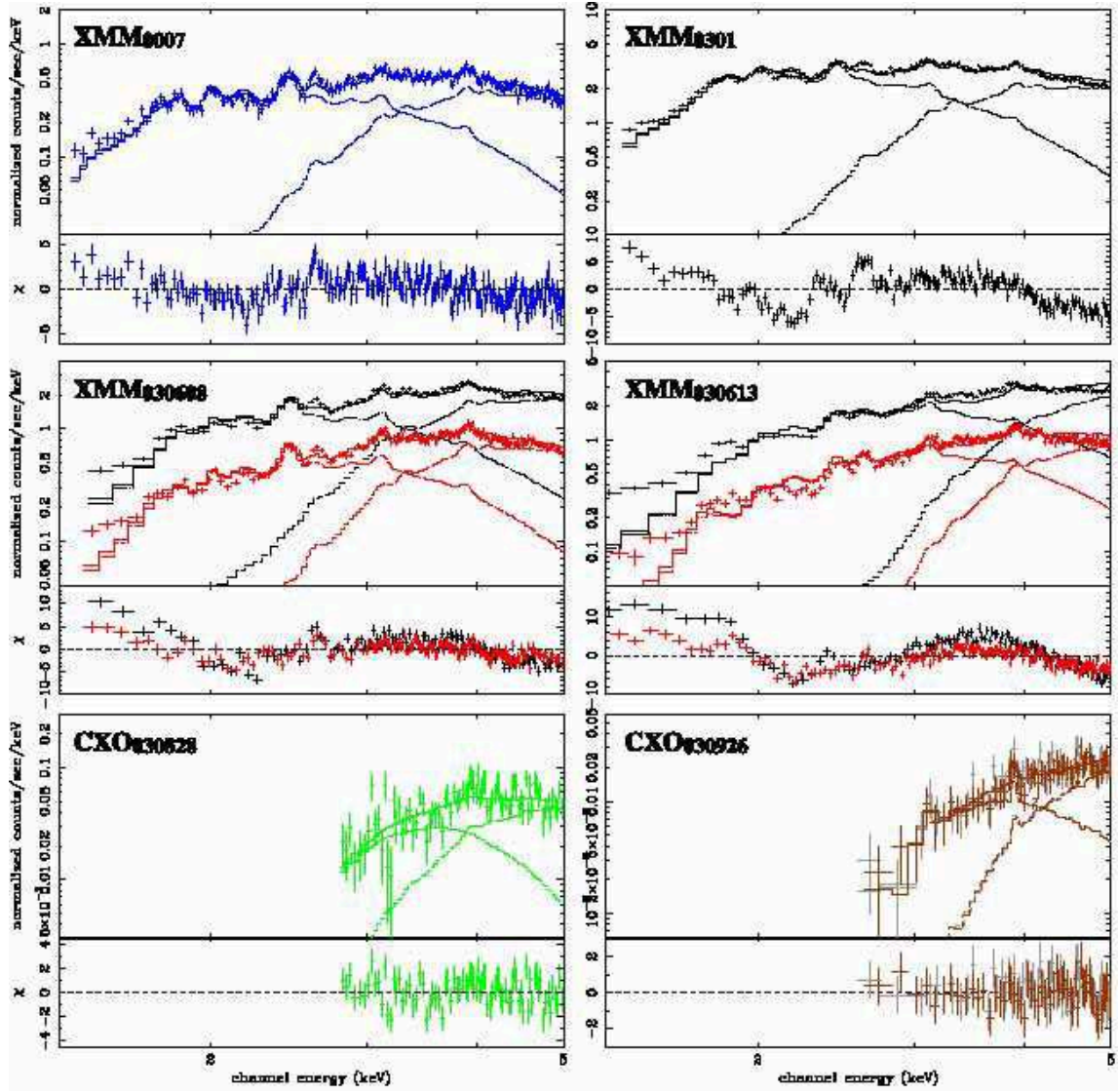}
\caption{Fits of the spectra below 5~keV.  
\label{fig:softfit}
}
\end{figure*}

\input{tab6}

Extrapolation of the model spectrum above 5~keV from the 
XMM$_{\rm 0007}$, XMM$_{\rm 0301}$, XMM$_{\rm 030608}$, XMM$_{\rm 030613}$, CXO$_{\rm 030828}$ and CXO$_{\rm 030926}$ data
to lower energies underestimates the observed emission at $E<$5~keV,
indicating the presence of additional cooler emitting material.
We thus modeled the emission below 5~keV by fixing the best-fit model spectrum derived in  
\S\ref{sec:ironkprofile}, and adding another 1T component 
with independent absorption. 
The results are given in Table~\ref{tbl:spec2tfit} and Figure~\ref{fig:softfit}.

These fits were not accepted at $>$90\% confidence
except the fit to the CXO$_{\rm 030926}$ spectra.
One reason for the poor quality of the fits is that the fixed hard component,
which reproduces emission above 5~keV,
somewhat overestimates the emission near 5~keV,
which might indicate we slightly underestimated the absorption to the hard component.
Another reason is that the best-fit models do not reproduce the strengths or locations of 
many emission lines, in particular \ion{S}{14}, which is generally emitted by plasma at \KT\  $>1$ keV.
This means that the spectra may require additional components with \KT\  between 1$-$3~keV.
The \NH\  values we derive from this 2 component modeling are very close to the ones from the 1T fitting in \S\ref{sec:varspecallfit}.

\section{Comparison with the Previous Observations Near X-ray Minima}
\label{sec:comppremin}

The weekly to daily monitoring of \etacar\  with \RXTE\  \citep{Corcoran2005} showed
significant cycle-to-cycle variability in flux and X-ray hardness.
Since the \RXTE\  data are contaminated by instrumental and cosmic background,
we compared our \XMM\  and \CHANDRA\  results with earlier spatially-resolved observations 
from \ASCA\  \citep{Tanaka1994}, \SAX\  \citep{Scarsi1993} and \ROSAT\  \citep{Trumper1984}.

Between phase $-$0.10 $\lesssim \phi_{orbit}\lesssim$ 0.10,
there were 4 \ASCA\  observations 
and 1 \SAX\  observation near the 1998 minimum,
and 1 \ROSAT\  observation during the 1992 minimum. 
These observations are summarized in Table~\ref{tbl:obslogsprecycle}. 
Because of the extended point responses (FWHM $\sim$a few arcminutes) of \ASCA\ 
and \SAX\ 
compared to \XMM, source spectra extracted from these observations used relatively 
large ($\sim 3'$) extraction regions \citep{Corcoran2000,Viotti2002}. 
This extraction area is about a factor of $\sim$20 larger than the area used in the \XMM\  analysis
(XMM SRC), and about a factor of $\sim$5000 larger than the area used in the \CHANDRA\  analysis
(CS SRC).
Because the neighborhood of \etacar\  is crowded with X-ray sources and diffuse emission, 
the \ASCA\ and \SAX\ 
spectra will suffer greater contamination from neighboring sources.

\begin{figure*}[t]
\epsscale{2.2}
\plottwo{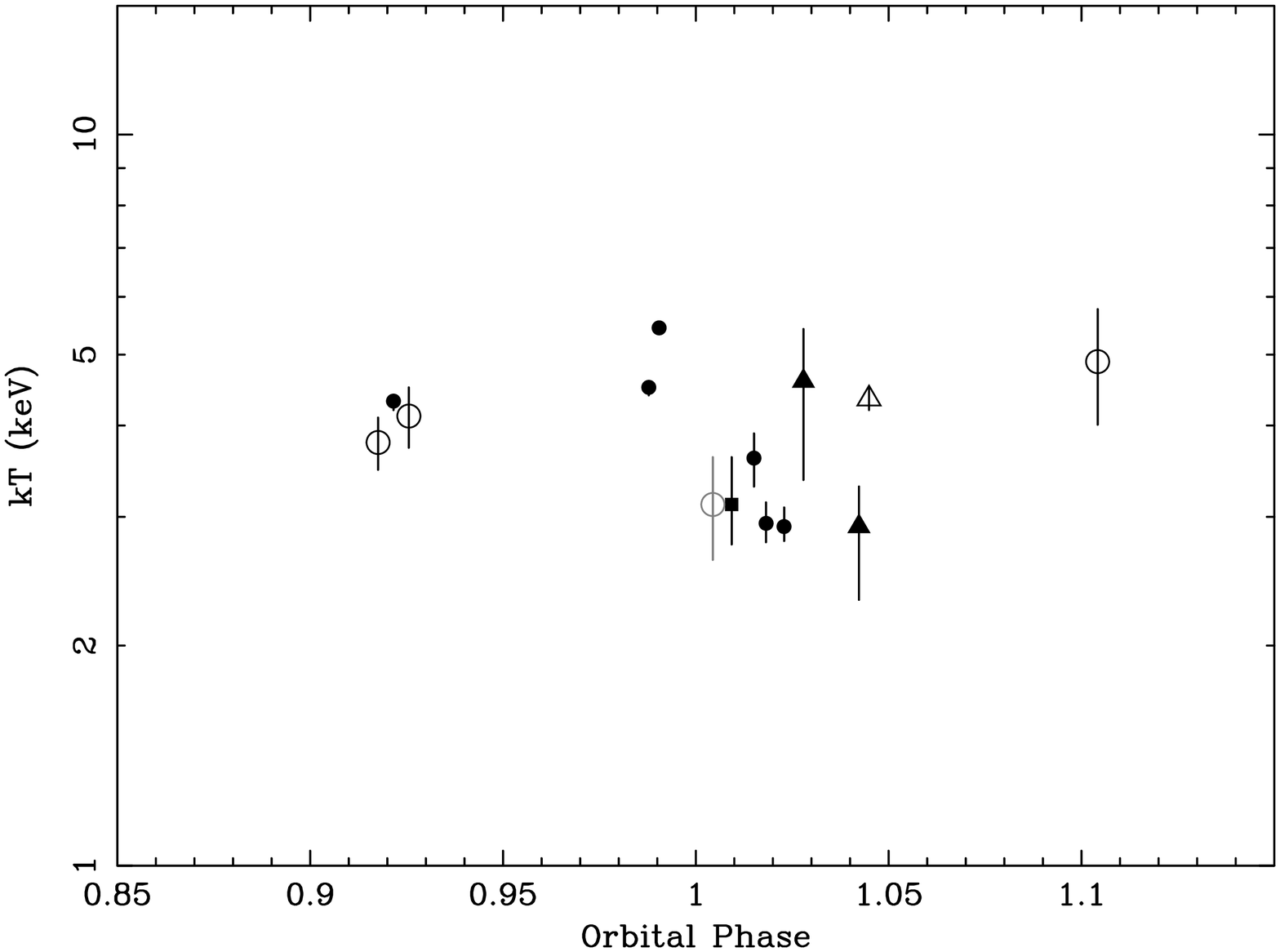}{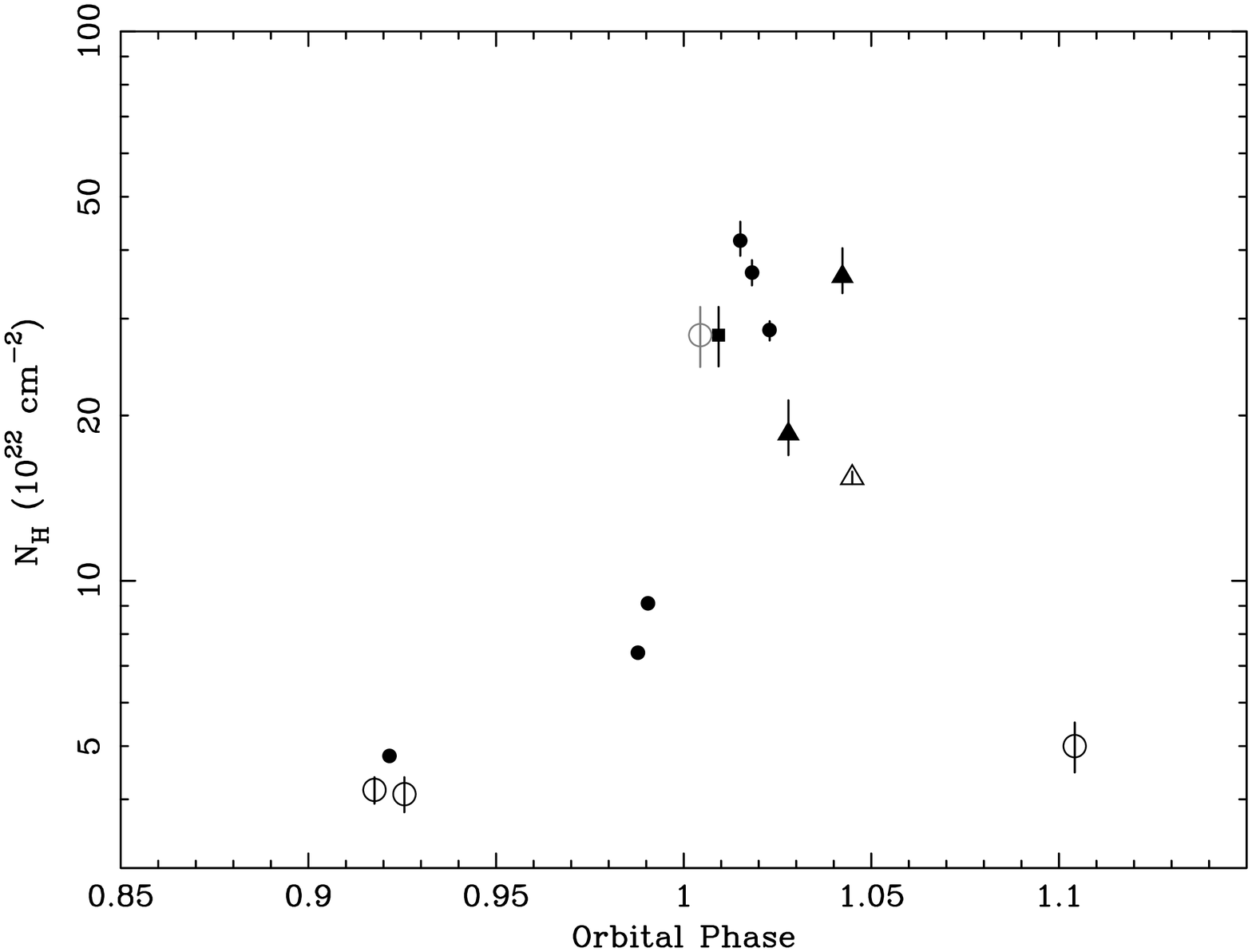}
\caption{Variation of the temperature and absorption column (\NHONET) vs. orbital phase.
The filled marks are from the 2003 cycle (circle: \XMM, triangle: \CHANDRA, square: simultaneous
fit between \XMM\  and \CHANDRA). 
The open marks are from the  1997/98 cycle
(circle: \ASCA, triangle: \SAX). The \ASCA\  spectrum during
the 2003 minimum at phase 1.004 is consistent with the \NH\  and \KT\  values from 
CXOXMM$_{0307}$, as shown.
\label{fig:abscorwearly}
}
\end{figure*}

\input{tab7}

\subsection{Comparison of the 1997/98 and 2003 Spectra Outside the Minimum}

Outside the minimum, contamination 
of the stellar source spectra in the \ASCA\  and \SAX\  observations
from the surrounding
sources, the CCE component, and the reflection from the Homunculus
Nebula is $\lesssim$10\% above $\sim$2.5~keV (see \S\ref{subsec:surcontami}).

\citet{Viotti2002} tried two models to fit their \SAX\  spectra: an absorbed single temperature bremsstrahlung model $+$ Gaussian for the spectrum
between 3$-$10~keV, and an absorbed two temperature model to fit
the entire spectrum between $\sim$0.1$-$10~keV. 
Both gave consistent results with each other for the central hard variable component.
On the other hand, in an analysis of \ASCA\  spectra obtained just prior to the 1997/98 minimum, \citet{Corcoran2000} fixed the emission from the outer ejecta using the model derived in \citet{CorcoranMF1998a}.
Though the coolest temperature they used to fit the \ASCA\ spectra (\KT\ $\sim$0.3~keV) is a bit lower than our adopted temperature (\KT\ $\sim$ 0.6~keV) and
the resulting difference in the outer ejecta flux is about a factor of three at $\sim$2~keV, this only produces a
$\sim$1\% uncertainty in the flux of the variable emission outside the 2003 minimum.
The results of their \SAX\ and \ASCA\ analyses, therefore, are comparable to the result
we derived from
our 1T fits to the entire \XMM\ and \CHANDRA\ spectra in \S\ref{sec:varspecallfit}.

The plasma temperatures in the 1997/98 observations were always around $\sim$4$-$5~keV,
which is similar to our results.
The derived values of \NH\  varied significantly with phase
(Figure~\ref{fig:abscorwearly}), reaching a maximum during the X-ray minimum, and declining thereafter. 
The variation in 2003 was similar to the \NH\ variation in 1997/98, 
except for the interval just at the end of the minimum.
Though \SAX\  and \CHANDRA\  observed \etacar\  at a similar orbital phase 
($\phi_{\rm SAX980318} \sim$0.045, $\phi_{\rm CXO030926} \sim$1.042),
the \NH\  measured with
\CHANDRA\  in 2003 is a factor of two larger than the value derived by \citet{Viotti2002} from the \SAX\  observations in 1998.
This is consistent with the behavior of the hardness ratio after the recovery as measured by \RXTE,
which was higher in the 2003 recovery than in the 1998 recovery \citep{Corcoran2005}.

\begin{figure*}
\epsscale{2.2}
\plottwo{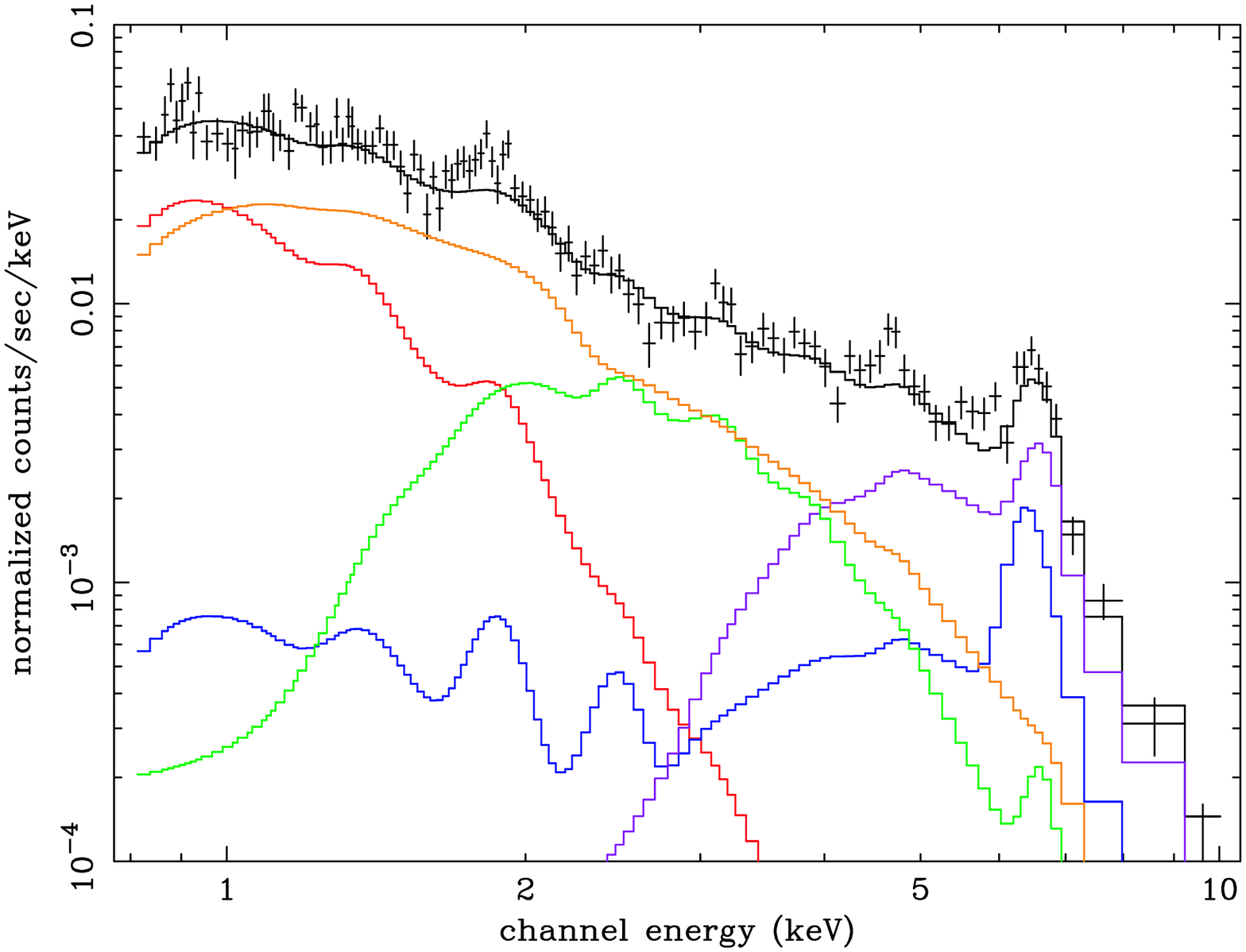}{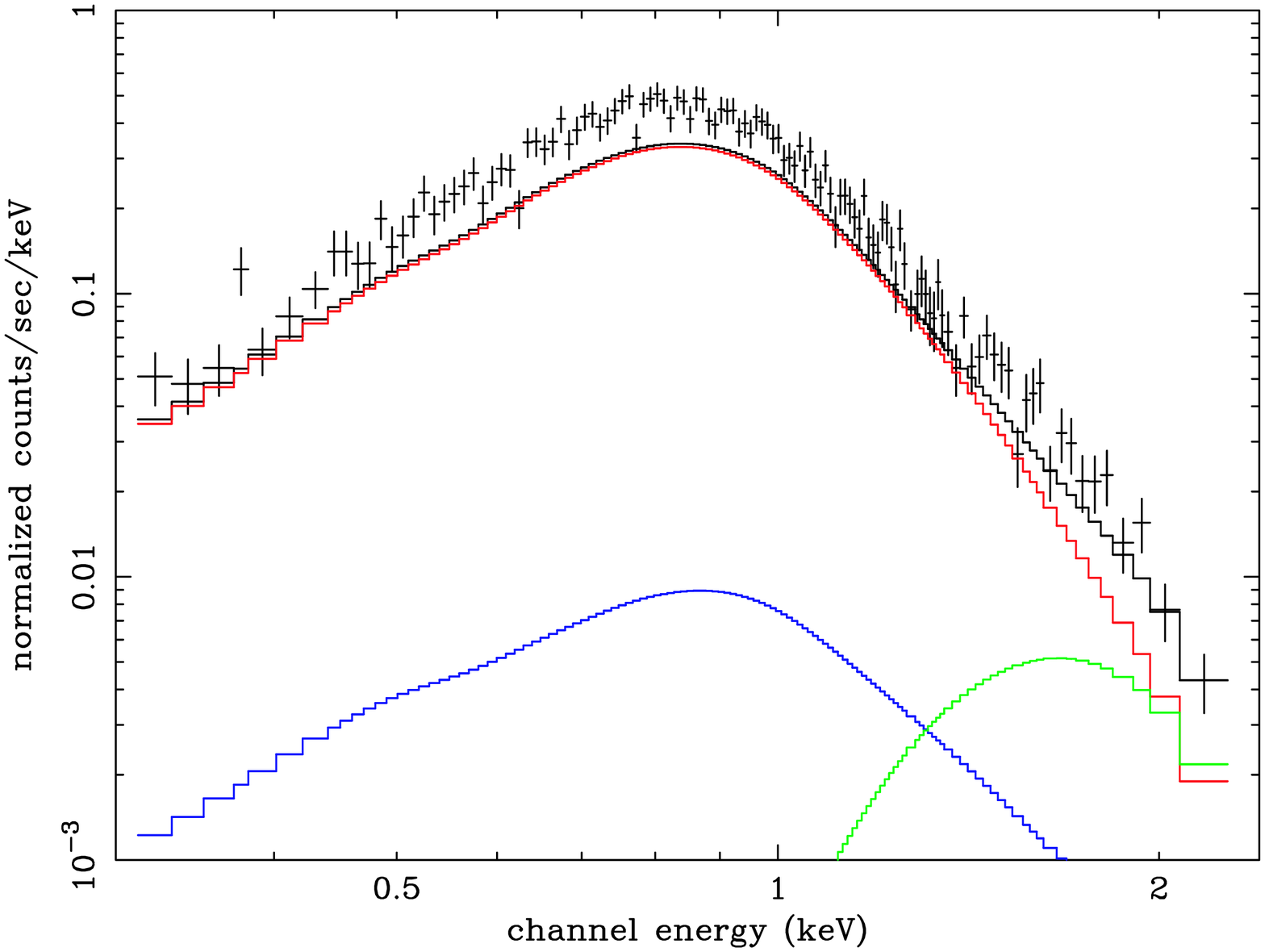}
\caption{{\it Left}: \ASCA\  GIS2+3 spectrum on 1997 Dec. 24. {\it Right}: \ROSAT\  PSPC spectrum
on 1992 Jun 12.  In each plot the spectrum of \etacar\ 
during the X-ray minimum in 2003 is shown 
({\it red}: the outer ejecta, {\it blue}: X-ray Homunculus Nebula, {\it green}: \centconst\  component,
 {\it purple}: variable emission in CXOXMM$_{\rm 0307}$; {\it orange}:
surrounding sources within 3\ARCMIN. The solid black line shows
the sum of these emission components.
\label{fig:ascaspec}
}
\end{figure*}

\subsection{Comparison of the 1997/98 and 2003 Spectra During the Minima}
\label{sec:minimumasca}
\label{subsec:surcontami}

In order to compare this dataset with the X-ray observations during the 2003 minimum,
we re-analyzed the ASCA$_{\rm 971224}$ GIS data.
We screened the ``revision2'' data taken from the HEASARC 
archive\footnote{http://heasarc.gsfc.nasa.gov/W3Browse/}
using {\it gisclean} and filtered it with the standard criteria,
which excludes data affected by the South Atlantic Anomaly, Earth occultation and 
high background in regions with low geomagnetic rigidities.
For both GIS2 and GIS3 detectors, 
we used a 3\ARCMIN\  radius circle centered on \etacar\  as the source region ``3\ARCMIN\  SRC",
and used as background emission from a 3\ARCMIN\  radius circle designated ``3\ARCMIN\  BGD''
centered to the north-east of \etacar\ (Figure~\ref{fig:obsimages})
where the soft diffuse X-ray emission from the Carina Nebula is apparently weak
\citep{Hamaguchi2006}.
We used the standard GIS response files, version 4.0 (g[23]v4\_0.rmf)
and generated ancillary response functions with {\it ascaarf} ver. 3.10.
We then merged both GIS2 and GIS3 spectra together to improve signal to noise.
The left panel of Figure~\ref{fig:ascaspec} shows the GIS spectrum from ASCA$_{\rm 971224}$. 

\begin{figure}[h]
\epsscale{1.0}
\plotone{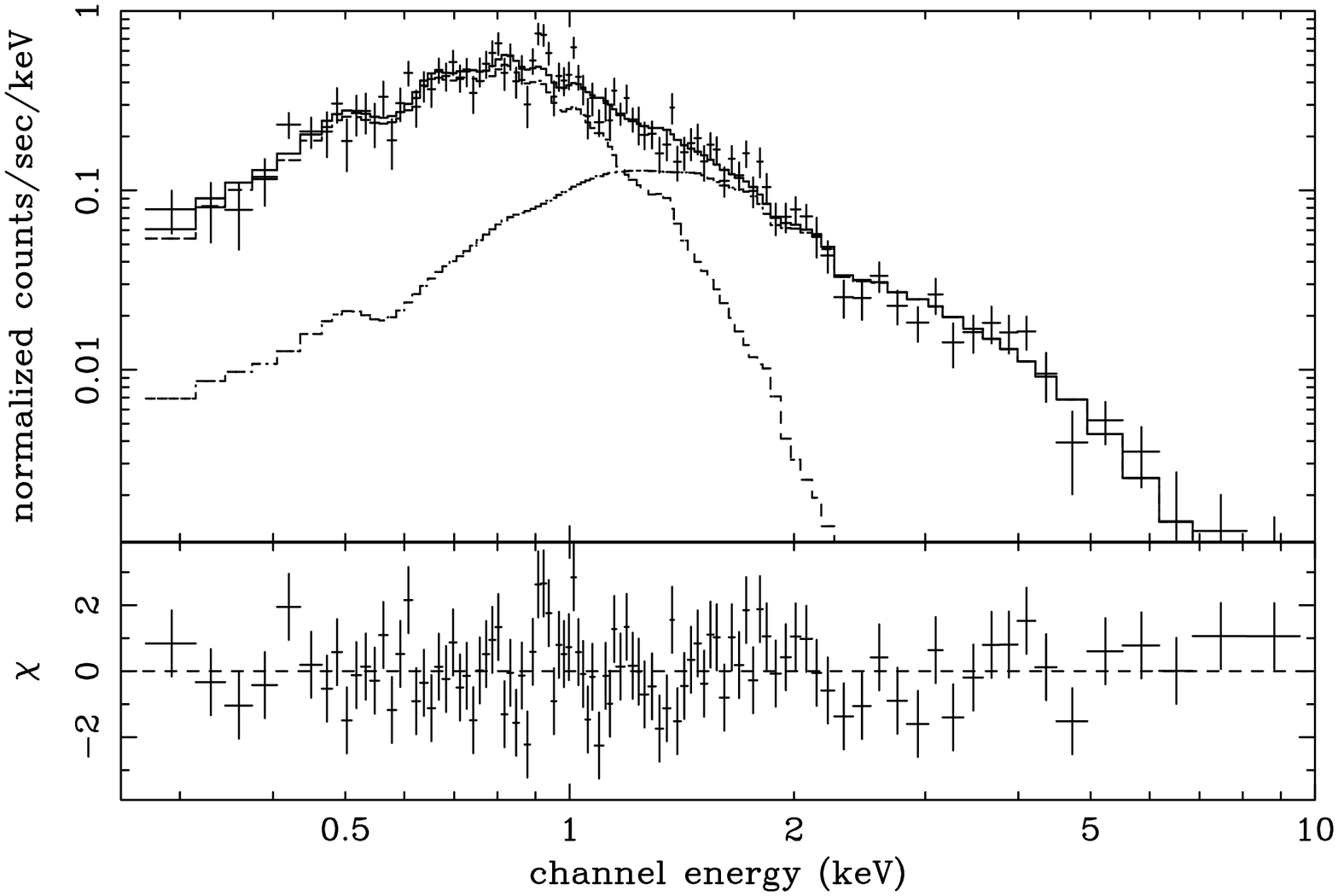}
\caption{\XMM\  MOS2 spectrum from XMM$_{\rm 030722}$ extracted from a 3\ARCMIN\  radius circle
excluding emission from \etacar.
The solid line shows the best-fit 2T MEKAL model with 
common absorption, while the broken lines show the individual fit components.
\label{fig:surroundingsource}
}
\end{figure}

Because the stellar source was faint during the minimum and the \ASCA\  point spread function is large, contamination by nearby X-ray sources is significant for this observation.
We therefore tried to estimate the  contamination from emission from sources within 3\ARCMIN\  of \etacar\  ``3\ARCMIN\  SRC" using XMM$_{\rm 030722}$
(when the central point source of \etacar\  was the weakest)
excluding the source region ``XMM SRC" used for the analysis of \etacar\  from the \XMM\  MOS2 data.
We used the same background region as 
we used to estimate the \ASCA\ background emission, ``3\ARCMIN\  BGD".
About 10\% of the emission from \etacar\  (which amounts to $\sim$20\% of the emission
from the surrounding sources even in XMM$_{\rm 030722}$) contaminates the ``3\ARCMIN\  SRC'' region.
We simulated the contamination due to \etacar\  and subtracted it from the \ASCA\
source spectrum.
The subtracted spectrum (Figure \ref{fig:surroundingsource}) does not show 
any evidence of the extremely strong nitrogen line at $\sim$0.5~keV from the outer ejecta, 
suggesting that contamination is well removed.

\input{tab8}

The resulting spectrum, after subtracting the simulated \etacar\  spectrum, shows weak emission 
lines at $\sim$0.9 and 1~keV from \ion{Ne}{9} and \ion{Ne}{10}, $\sim$0.8~keV possibly from Fe,
and a marginal excess around $\sim$3--4~keV from Ca.
This spectrum can be fit either by an optically thin, thermal (MEKAL) 
model with 2-temperature components 
or by a cool component plus a power-law component for the hard emission
at slightly below the 90\% confidence,
assuming common absorption for both components (Table \ref{tbl:specsurround}).
In either model, the average absorption to these sources is consistent with the interstellar value
\citep[$\sim 3\times 10^{21}$\UNITNH, see \S2.2 of ][]{Leutenegger2003}.
As the spectrum does not have strong emission lines, the elemental abundance of the 2T model 
is small ($\sim$0.04~solar), while the abundances are not constrained by the 1T plus 
power-law model.
We thus fixed the abundances at 0.3~solar, a typical value
for stellar X-ray emission obtained from low resolution CCD spectra
(e.g. OB stars, \citealt{Kitamoto1996,Kitamoto2000}; low-mass MSs, \citealt{Tagliaferri1997}; low-mass PMSs, \citealt{Yamauchi1996,Kamata1997,Tsuboi1998}.)\footnote{
High resolution grating spectra with \CHANDRA\ and \XMM\
showed complex line profiles especially from OB stars,
and therefore the abundance obtained from low resolution spectra may not
reflect real elemental abundance.
\citep[e.g., ][]{Cassinelli2001,Kahn2001,Miller2002}
}
The observed flux is $\sim$3.2$\times$10$^{-12}$~\UNITFLUX\ (0.5--10~keV).

The ``3\ARCMIN\  SRC'' region includes at least 55 \CHANDRA\  X-ray point sources \citep{Evans2003}:
6 OB stars and 49 unidentified, weak sources that might be low mass pre-main-sequence stars.
The combined absorption corrected flux from these sources calculated from Table 3 and 4 
in \citet{Evans2003} is 1.3$\times$10$^{-12}$~\UNITFLUX\  (0.5$-$2.04~keV).
There is also diffuse emission within the source region which is also calculated from 
\citet{Evans2003} to have a flux of $\sim$5.7$\times$10$^{-12}$~\UNITFLUX.
The total absorption corrected flux, $\sim$7$\times$10$^{-12}$~\UNITFLUX, is roughly consistent 
with our result 8.2$\times$10$^{-12}$~\UNITFLUX\  (0.5$-$2.04 keV, unabsorbed).
Moreover, the combined absorption corrected fluxes of the point sources are as large as the flux 
from the hard component (1.7$\times$10$^{-12}$~\UNITFLUX), and the flux from the 
diffuse emission is as large as the soft component, 6.5$\times$10$^{-12}$~\UNITFLUX, 
in the $0.5-2.0$ keV band.
As seen in Figure~\ref{fig:obsimages}, many point sources are detectable at intermediate energies,
where the hard component is dominant,
but they are not clear in the soft band (except for HDE~303308).
The hard component probably represents emission from point sources with the soft 
component dominated by diffuse emission and HDE~303308.

\label{subsec:ascaana}

We have compared the \ASCA\ spectrum to a model 
derived from our fits to the \CHANDRA\  and \XMM\  spectra including emission from the outer ejecta,
from the Homunculus Nebula, from the \centconst\  component
and the variable model which fits the spectrum obtained from the
CXOXMM$_{\rm 0307}$ observation, along with emission from the surrounding sources
(diffuse emission and point sources) which fall within the 3\ARCMIN\  extraction circle.
The \ASCA\  spectrum agrees well with this combined model.
Assuming that emission from the outer ejecta, X-ray Homunculus Nebula and the surrounding sources did
not vary between 1997 and 2003 and that spectral shapes of the \centconst\  and variable components
did not change dramatically, the flux of both \centconst\  and variable components are
the same within 50\%.

\subsection{Comparison with the 1992 Minimum}

There is one \ROSAT\  PSPC observation of \etacar\  during the minimum in mid-1992.
\citet{Corcoran1995} used this observation to show that the X-ray emission in the hard \ROSAT\  band 
($E >$1.6~keV) decreased by a factor of 2 at that time. \ROSAT\  however had very little effective area above 2 keV so that the bulk of the variable emission was not observable by \ROSAT.

\citet{Corcoran1995} extracted a spectrum from a 1\FARCM85 radius circle excluding 
a 30\ARCSEC\  radius circle around the nearby bright star HDE~303308.
To minimize contamination from surrounding sources,
we re-extracted the spectrum using an elliptical source region of 65\ARCSEC\  $\times$ 47\FARCS5,
which includes emission from the outer ejecta but which excludes HDE~303308 and 
point and diffuse sources around \etacar.
We took the same background region as used in our analysis of ASCA$_{\rm 971224}$.
In analyzing the PSPC spectrum, we used the standard response file pspcb\_gain2\_256.rmf
from the \ROSAT\ calibration databbase, 
but we generated the ancillary response file using the pcarf v. 2.1.3 as appropriate for this observation.

The PSPC spectrum is shown in the right panel of Figure~\ref{fig:ascaspec}.
We have overlaid all the spectral components of \etacar\ on this spectrum as we did for
the ASCA$_{\rm 971224}$ spectrum, except for the component due to the surrounding sources,
whose contribution would be small in the \ROSAT\  spectrum due to the smaller PSPC extraction region.
Emission between 0.5$-$1~keV, where the outer ejecta component is dominant, is about a factor of 30\%
larger than the model.
This difference may be due to calibration uncertainties between \ROSAT\  and \XMM, 
or it might be produced by another variable source within the \ROSAT\  extraction area.
Though the \XMM\  model of the \centconst\  component only contributes to the few 
highest channels of the PSPC detector,
this component is consistent with the \ROSAT\  spectrum.  Thus it appears that \ROSAT\  did not detect the stellar component at all during the 1992 minimum.

\section{Discussion}

\subsection{Emission from the Variable Component}

\RXTE\  clearly showed that the variable emission is periodic
and leaves little doubt that the emission is produced by the WWC in a binary system \citep{Corcoran2005},
and if so, that the X-ray minimum
occurs near periastron passage.
Our results further strengthen this conclusion.
The temperature of the hottest plasma, $\sim$3~keV, was mostly unchanged through the cycle,
consistent with the collision of stellar winds at terminal velocities \citep[e.g., see ][]{Ishibashi1999}.
Furthermore, the spectral shape varied in a periodic way, as well.
However, the \NH\ measured just after the end of the X-ray minimum was about a factor of two 
larger during the 2003 cycle than it was during the 1998 cycle.

Most of the light curves derived from the \CHANDRA\  and \XMM\  data showed no short-term 
variations, suggesting that the X-ray plasma is produced by a steady source and is probably 
larger than $V_{shock}\times t_{exp}$, where $V_{shock}$ is the pre-shock velocity and
$t_{exp}$ is the exposure time of the observation.
For $V_{shock}\sim 3000$ \UNITVEL\  
(corresponding to shock temperatures of $\sim 3-4$ keV) and exposure times of $\sim10$ ksec (corresponding roughly to the \XMM\  observing time), 
the size of the emitting region is probably $\gtrsim 0.2$ AU, and perhaps considerably larger.  
The \RXTE\  lightcurve also shows clear variations in observations separated by as little as 1 day, 
suggesting that the size of the emitting region is
$0.2 \lesssim r \lesssim 2$ AU.
Although the orbital elements are currently uncertain, the stellar separation at 
periastron may be as little as 1.5 AU \citep{Corcoran2001b}, so that the size of the emitting region 
may be comparable to the stellar separation at periastron passage.

Spectra outside the 2003 minimum require at least 2 temperatures,
\KT\  $\sim$3~keV and 1--1.5~keV, consistent with the analysis of a 
\CHANDRA\  grating spectrum by \citet{Corcoran2001a}.
The failure to fit most of the spectra by 2 temperature models in 
\S\ref{sec:softbandfit} is perhaps an indication of the presence of more than 2 
temperatures.
This is consistent with a hydrodynamic simulation of the WWC plasma \citep{Pittard1997},
in which the plasma temperature is the highest near the stagnation point 
where the winds collide head-on, and cooler farther along the bow-shock.
The excess at the blue side of the \ion{Fe}{25} line discussed in \S \ref{sec:ironkprofile} may represent streams from the stagnation point flowing outward near the companion's wind terminal velocity.

The \RXTE\  light curves also displayed quasi-periodic X-ray ``flares", which 
correspond to hardness ratio maxima \citep{Corcoran1997,Corcoran2005}.
XMM$_{\rm 030608}$ occurred near the bottom of one of these flares, 
while XMM$_{\rm 030613}$ occurred near a flare peak.
A comparison of these spectra in Figure~\ref{fig:specnorm} shows that
the hard band flux went up by a factor of two,
while the  soft band flux below $\sim$3~keV did not change except that
emission lines are apparently weaker in XMM$_{\rm 030613}$.
This is consistent with an increase of the hardness ratio measured with \RXTE.
This indicates either a slight increase in \KT\  and/or \EM\  for the hottest plasma,
possibly caused by increase of the density of the high-temperature gas near the stagnation point, as discussed for a similar event in the previous cycle observed with \ASCA\ 
\citep{Corcoran2000}.
The emission at $E\lesssim 3$ keV, on the other hand, did not vary strongly
between those observations.
One possibility is that hard X-ray emission from the highest-temperature plasma 
at the stagnation 
point of the shock cone varied strongly due to density fluctuations of the primary or secondary wind, 
with the soft X-ray emission coming from a larger area which averages out such fluctuations.
Another possibility is that the flares are related to instabilities in the WWC plasma.

\begin{figure*}
\epsscale{2.1}
\plotone{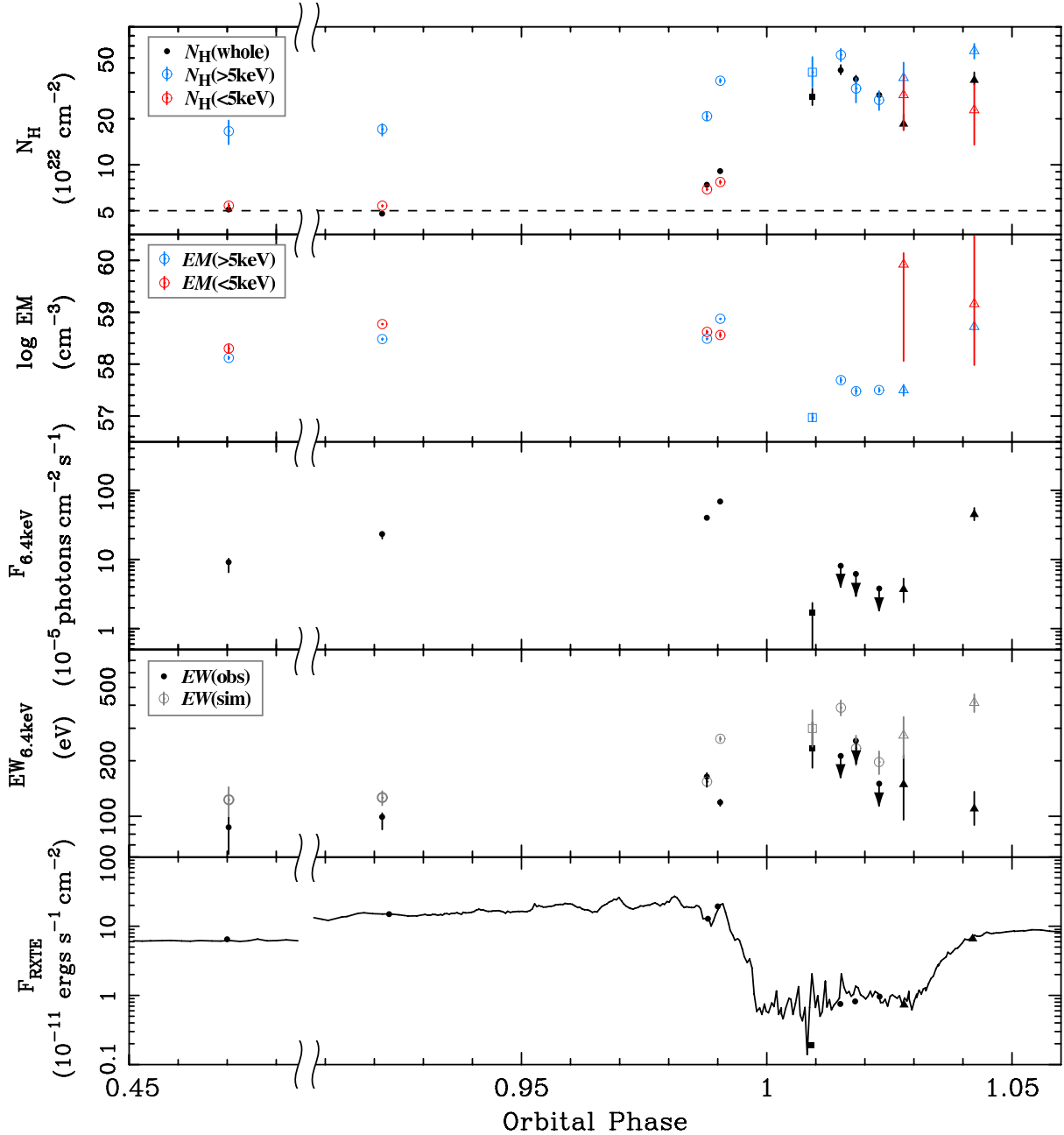}
\caption{Variation of \NH, \EM, intensity and EW of the Fe fluorescence line and 
\RXTE\  flux in the 2003 cycle.
The \XMM\  results are shown as circles, \CHANDRA\  results as triangles, while squares show results derived from a simultaneous
fit between \XMM\  and \CHANDRA.
The dash line in the top panel represents our estimate of the absorption beyond the central source.
\label{fig:nhnfensoft}
}
\end{figure*}

\subsection{The Absorber around the WWC Plasma}
\label{subsec:abswwcplasma}

Before the 2003 minimum, \NH\  measured from our fits to the spectra above 5~keV (hereafter \NHHOT) 
was always about a factor of four larger than \NH\  measured from the spectra below 5~keV
(\NHCOOL; Figure~\ref{fig:nhnfensoft}).
\NHHOT\  mainly represents Fe absorption, while \NHCOOL\  represents absorption by lighter elements
\citep{Morrison1983}.
Though this discrepancy can be produced by an Fe overabundance in the \etacar\  ejecta, 
the elemental abundances around the stellar source are 
near solar except for He and N, which are overabundant, and O and C, which are strongly depleted \citep{Hillier2001,VernerE2005a,Davidson1984}.
Abundance anomalies of He, N, O and C would not change the amount of absorption above $\sim$1~keV.
This means that the soft emission does not suffer as strong absorption as
the hard emission before the minimum, and hence this supports the idea that
the variable emission originates in multiple-temperature plasmas with various absorptions.
This is consistent with the idea that the hot emission originates from a  highly absorbed region 
near the apex of the shock cone, with the lower-energy emission arising from a region farther 
along the shock interface which suffers less absorption.
Neither \NHHOT\  nor \NHCOOL\  varied significantly between XMM$_{\rm 0007}$
at $\phi\sim$0.47 (close to apastron) and XMM$_{\rm 0301}$ at $\phi\sim$0.92,
when the stellar separation changed by about a factor of 3.
The value of \NHCOOL\  before the minimum 
($\sim$5$\times$10$^{22}$~\UNITNH) may 
be absorption beyond the wind of the primary.
The higher value of \NHHOT\  suggests that extra absorbing material must be located near 
the hottest plasma.

After XMM$_{\rm 0301}$ and through the onset of the 2003 minimum, 
\NHHOT\  
and the absorption column derived from fitting the spectra over the entire $1-10$ keV band 
(\NHONET, see Figure~\ref{fig:abscorwearly})
increased to 3$-$5$\times$10$^{23}$~\UNITNH.
As shown in Figure \ref{fig:abscorwearly} 
the observed maximum column density occurred near the mid-point of the X-ray minimum, and the column densities declined thereafter towards the end of the minimum. 
During the X-ray minimum,
\NHCOOL\  was unmeasurable because of the near total suppression of the low energy flux.
This could mean
that the \EM\ of the cool component decreased during the minimum, or that
the absorption to the cool plasma also increased on the line of sight.
Nevertheless, the \NHHOT\ 
confirms that the X-ray minimum is closely associated with an increase in the amount of absorption along the line of sight.
This suggests that the WWC plasma entered into, or was hidden behind, the densest part 
of wind from the primary star.
The behavior of the absorption towards the hard component, \NHHOT, is in interesting contrast to the behavior of \NHONET.
\NHHOT\  had already increased to $\sim$4$\times$10$^{23}$~\UNITNH\ 
by XMM$_{\rm 030613}$ (i.e. prior to the start of the X-ray minimum) 
and it did not change strongly after that.
This could be explained if the WWC region was bent, perhaps by the Coriolis force,
causing the stagnation point to enter into the densest part of the primary's wind earlier than 
the downstream material.

The column density did not decline until some time after the 2003 minimum ended.
This confirms that the column density variation is not completely synchronous 
with the X-ray minimum.
Conjunction probably occurred during the interval when \NH\  was near its maximum observed value, in the phase interval $0.99<\phi<1.05$.
We note that the \NH\  measured in 2003 ($\phi$=1.042) is a factor of two larger 
than the \NH\  measured in 1998 ($\phi$=0.045).
\citet[ see their Figure~2c]{Davidson2005} also noted that the absorption component in the P-$\eta$($\lambda$9017) line in 2003.72 ($\phi\approx 1.0$) was stronger than in 
1998.21 ($\phi\approx 0.0$), 
which might also indicate that the amount of absorbing material in the line of sight 
near the start of the X-ray minimum is increasing with time.
However they also note that the equivalent widths of the H-$\alpha$ emission lines were lower just after the end of the 2003 minimum.
This might suggest a distribution of wind material which has preferentially 
grown denser in the line of sight than perpendicular to the line of sight in the 1998$-$2003 interval.

\begin{figure*}
\epsscale{2.1}
\plotone{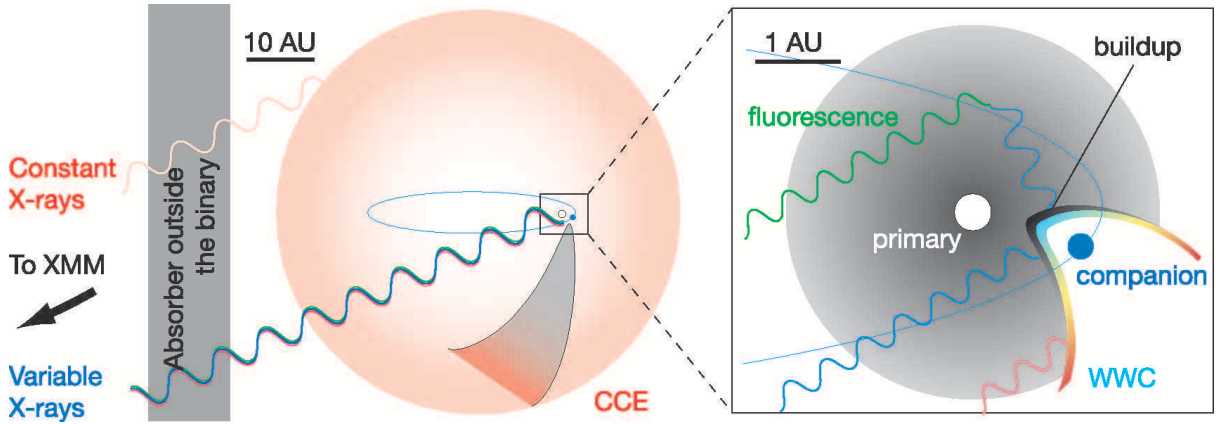}
\caption{Cartoon of the \etacar\  system.
Winds from the primary star and the companion collide together, to form the 
wind-wind collision (WWC) region which has a conical shape. Behind the WWC boundary is a low density region filled with the wind of the companion star.
The apex where the winds collide head on has hotter gas than the downstream material
and also accumulates a buildup of cool primary wind material in front.
We view the X-ray emission from the WWC plasma through 
the cool material at the apex and/or the undisturbed primary wind, and the absorber which lies far outside the binary system.  Iron K fluorescence emission is produced by scattering in the primary wind.
One possibility for the \centconst, where the \centconst\  is produced in outer shock, is shown.
Emission from the \centconst\ 
only passes through the outer absorber beyond the binary system. 
\label{fig:cartoonetacar}
}
\end{figure*}

The equivalent width (EW) of the Fe K fluorescence line varied between 100$-$200~eV,
which is far smaller than the EW ($\sim$1.5~keV) produced by scattering X-rays from the 
Homunculus Nebula \citep{Corcoran2004}.
This suggests that the X-ray emission in all the observations comes 
directly from the WWC plasma, 
even during the X-ray minimum, with only a small scattered component (see Figure~\ref{fig:cartoonetacar}).
From \citet{Inoue1985}, the Fe K fluorescence line EW produced by scattering from a spherically-symmetric medium 
around an X-ray source is EW $\sim$7.5$\times$\NHHOT/10$^{22}$~eV, 
where we have multiplied Inoue's coefficient by 0.75, the flux ratio in the $7.1-9$ keV band
between a \KT\  $\sim$3.3~keV thermal emission component appropriate to the \etacar\  spectra and the 
$\Gamma=1.1$ power-law emission component assumed by \citet{Inoue1985}.
In Figure~\ref{fig:nhnfensoft}, we plot the expected Fe-K fluorescence line EWs as open 
symbols for a spherically-symmetric scattering medium.  
The differences between the expected and observed equivalent widths seem to be larger 
when the X-ray flux goes up and smaller when it goes down
(XMM$_{\rm 030608}$, XMM$_{\rm 030613}$ and CXO$_{\rm 030926}$),
which may suggest that the variation of the fluorescence line is delayed by light travel time.

The observed EWs are $\sim$25\% smaller than the expected EWs even during the XMM$_{\rm 0007}$ ($\phi$=0.470) 
and XMM$_{\rm 0301}$ ($\phi$=0.923) observations when the direct X-ray emission did not vary strongly.
This suggests that the scattering material is not spherically symmetric 
as assumed in \citet{Inoue1985},
and that there is a deficit of scattering material off the line of sight.    
This deficit is probably produced by the wind-wind collision, since the region on the companion side of the shock has a much lower density than on the primary star's side.  In order to reduce the fluorescent Fe emission by 25\%, the  full opening angle of the wind shock cone would need to be $\sim$120\DEGREE. 
This is consistent with the measurement of the half-opening angle of $50-120$\DEGREE\  by \citet{Pittard2002}.

\subsection{The Nature of the X-ray Minimum}

The 2003 X-ray minimum can be described as an apparent decrease in \EM, along with an increase in \NH. 
However, a column density of \NH\ $\sim$5$\times10^{24}~$\UNITNH\  is needed
to reduce the emission at 10~keV by two orders of magnitudes, 
and the observed maximum column density (\NH\ $\sim 5\times 10^{23}$~\UNITNH) is much lower than this.    The \ASCA\  observations during the 1997/98 minimum \citep{Corcoran2000} also appeared consistent with an apparent decline in \EM

Two mechanisms could produce the observed decrease in \EM\
One possibility is an ``eclipse model", namely that emitting region is almost totally, but not completely, obscured by an optically thick absorber 
with \NH$\gtrsim10^{25}$~\UNITNH, 
and we only see the least absorbed residual emission during the minimum.
\citet{Ishibashi1999} and \citet{Corcoran2001b} tested the eclipse model using the
RXTE light curve of \etacar\  and showed that the duration of the X-ray minimum ($\sim 3$ months) at high energies ($E>5$ keV) was difficult to explain for
a spherically symmetric wind from \etacar\  if the mass loss rate was constant at $\sim 10^{-4}$ M$_{\odot}$ yr$^{-1}$. The eclipse model seems to require that the wind is not spherically symmetric, or that the absorption to the hard X-ray emitting region was enhanced for an extended interval near periastron passage.
It also suggests that a portion of the hard emission is visible during the minimum so that
the size of the hottest region is greater than the size of the occulting region.

Another possibility is that the emissivity of the WWC itself fades during the minimum, i.e. that there is a decline in the amount of material hot enough to generate $\sim5$ keV X-rays.
Some WWC models predict a change in the X-ray emissivity of the shock
around periastron. For example, ``radiative braking''
\citep{Gayley1997} in which UV photons from \etacar\  decelerate
the companion's wind at phases near periastron could reduce the intrinsic X-ray emission from the wind-wind collision at $E>5$ keV. \citet{Davidson2002} suggested that strong instabilities near periastron might cause the shocked gas to radiatively cool, reducing the emission at high energies.
\citet{Soker2005} and \citet{Akashi2006} suggested that Bondi-Hoyle accretion of the primary star's wind by the companion star near periastron
could suppress the companion's wind and thus reduce the emission from the colliding wind shock for a brief period.
In optical observations,
the \ion{He}{2} $\lambda 4687$ emission line, which is believed to arise near the WWC shock,
decreased in concert with the X-ray drop \citep{Steiner2004}. 
\citet{Martin2006} argued
 that the disappearance of this line in a direct view of the star, and in a reflected polar view \citep[the so-called ``FOS4'' position, ][]{Stahl2005}
indicated a waning of the shocked gas during the minimum.

Below, we examine some of the behaviors of the electron temperature, iron line profile, iron fluorescence line variation and lack of variation at E $>$ 7~keV during the minimum from the 2003 X-ray spectra to try to decide between these two mechanisms.

\subsubsection{Electron Temperature}
The electron temperature of the hottest material
did not change significantly during the 2003 minimum, as the normalized spectra at the bottom of 
Figure \ref{fig:specnorm} clearly show.
The constancy of the electron temperature of the hottest material argues against radiative braking, shock instability or accretion models which predict that the plasma temperature should decrease.
\citet{Akashi2006}'s model in particular predicts complete shutdown of the WWC activity during periastron passage,
and therefore could only be consistent with a constant electron temperature if an alternative X-ray emission mechanism
during the minimum could produce plasma at nearly the same temperature as the WWC emission.
If the shock actually disappears during the minimum, then
the constancy of the high-energy X-ray continuum  requires
that a small part of the 3000~\UNITVEL\  secondary wind continues
to produce a shock by encountering material directly at normal incidence.

\subsubsection{The Iron Line Profile}
The variable excess emission on the red side of the \ion{Fe}{25} line can be explained
by non-equilibrium ionization (NEI) effects in the high-temperature plasma, and
variations in the strength of this feature means that the ion temperature of some of the high-temperature 
plasma in the colliding wind shock changed during the minimum.
This might favor mechanisms in which the intrinsic emission from the shock declines with time
near periastron, though the NEI effect itself does not reduce emissivity of the continuum.
The observed excess at $\sim$6.55~keV is consistent with
emission lines of around \ion{Fe}{21}, present at
an ionization time scale $\tau = nt \sim 10^{10-11}$~\UNITIONTIME\ 
in the $\sim$3.3~keV plasma,
where $n$ is the gas density and $t$ is the time after the gas starts heating.
Since the gas density $n \propto D^{-2}$, it should be larger by
more than two orders of magnitude at periastron than at apastron
\citep[using the orbital elements of][]{Corcoran2001b}.  Thus near periastron the
plasma should reach ionization equilibrium  
much faster \citep[$\sim$15 min,][]{Soker2005} than at apastron.
Perhaps near periastron, both the cooling time-scale and the equilibrium time-scale decrease relative to their values at apastron, but the cooling time-scale decreases faster.
This means that near periastron the fraction of non-equilibrium gas to gas in equilibrium is larger than it is near apastron.
However,  since the radiative cooling time-scale has the same dependence on gas density 
as the ionization equilibrium time-scale ($\propto n^{-1}$), an additional cooling mechanism (such as adiabatic expansion or conduction) needs to be effective near periastron in order to decrease the cooling time-scale faster than the equilibrium time-scale.

\subsubsection{Iron Fluorescence Line Variations}

As noted in the previous section,
the EWs of the Fe K fluorescence line stayed below $\sim$200~eV during the 2003 minimum,
and therefore the observed X-ray emission is direct emission, with only a small scattered component.
The Fe fluorescence line flux decreased by a factor of $\sim$50 during the 2003 minimum
(Figure~\ref{fig:nhnfensoft}).
This could  be produced by reducing the number of $E >$7.1~keV X-ray photons from the source 
seen by the fluorescing region, or by reducing the number of photons from the fluorescing
region which reach the observer.
If we assume that the fluorescing region extends far from the stars then during the 2003 minimum either
the X-ray source is nearly entirely covered by obscuring material along the line of sight to the fluorescing
region, or the intrinsic X-ray emissivity has declined so that fewer ionizing photons reach the fluorescing
medium.
This picture is consistent with a similar variation of \ion{He}{2} $\lambda$4687 emission
between the direct emission from the central WWC region and polar emission scattered from 
the SE lobe \citep{Stahl2005,Martin2006}.

\subsubsection{The Lack of Variation at $E>7$ keV During the Minimum}

The hard band X-ray emission above $\sim$7~keV reached its lowest value after the 2003 minimum started,
recovered by a factor of 3 after XMM$_{\rm 030722}$ and did not vary for at least $\sim$3 weeks
from XMM$_{\rm 030802}$ to CXO$_{\rm 030828}$.
Since \RXTE\  light curves showed a similar trend in the previous cycle,
the time variation seems to be phase repeatable.
The observed X-ray flux after XMM$_{\rm 030722}$ was $\sim$5\% of the flux at 
XMM$_{\rm 030613}$.
It is very difficult for an eclipse of the X-ray emitting region by an opaque disk \citep[similar to that proposed by ][for example]{Ishibashi1999},
to keep the observed flux constant for so long a time, 
since the stellar motion would be so fast near periastron that the emitting region moves beyond the occulter 
on $\sim$day timescales, causing a short eclipse and a rapid rise in X-ray brightness.
Perhaps a ``leaky absorber'' eclipse, in which the absorbing medium is very extended and composed of optically thick clumps, 
could better describe the behavior of the observed flux during the minimum.  Models in which the shock cools may not be consistent with the constancy of the $E>7$ keV emission either, 
unless some residual emission at high energies could be generated after the shock cools.

\subsection{The Origin of the CCE Component}

The \centconst\  component did not vary within the two months during the
2003 minimum when it could be observed.
It apparently had the same spectrum during the 1998 minimum 
(as shown by our analysis of ASCA$_{\rm 971224}$), and did not show any enhanced emission during the minimum in 1992.
This suggests that the X-ray source is stable on time-scales of 5$-$10~years.
The \CHANDRA\  image during the 2003 minimum restricts the size of the emitting plasma 
to $\lesssim 1''$, constraining the maximum size of the \centconst\  at $\lesssim$2000$\mbox{AU}$.
The column density to this component, \NH\  $\sim$5$\times$10$^{22}$~\UNITNH, is smaller 
than the column density to the variable emission component around the 2003 minimum,
but each component shows nearly the same column density in January 2003 
(before the 2003 minimum) and near apastron.
After correcting for absorption, the X-ray luminosity of the \centconst\  is $\sim$10$^{34}$~\UNITLUMI.
This suggests that a stable luminous source is present 
outside of the absorbing material which lies in front of the colliding-wind source at periastron.

One possibility is that the \centconst\  component is produced within the wind of 
the companion star and/or \etacar\  due to the inherent instabilities of their radiatively-driven winds.
The X-ray properties of the \centconst\  component -- 
\KT\  $\sim$1~keV with little apparent X-ray variation -- resemble the X-ray properties
of normal OB stars \citep[e.g.~][]{Berghoefer1997,Corcoran1993}.
For OB stars $\log$ \LX$/$\Lbol$\approx -6$ to $-8$ \citep{Berghoefer1997}, 
so that the X-ray luminosity of the \centconst\  corresponds to a bolometric luminosity
\Lbol\  $\sim 10^{40-42}$~\UNITLUMI.
This is somewhat larger than the luminosity of the companion according to 
\citet[][]{VernerE2005a} 
(\Lbol\  $\sim 4\times 10^{39}$~\UNITLUMI), 
but near the luminosity of \etacar\ 
\citep[\Lbol\  $\sim 2\times 10^{40}$~\UNITLUMI,][]{Hillier2001}.

Alternatively, the \centconst\  emission could be produced
in ambient gas shocked by a strong outflow from the stellar winds (see Figure~\ref{fig:cartoonetacar}).
The maximum size of the \centconst\  ($r \lesssim 2000\mbox{AU}$)
means that it is inside the Little Homunculus nebula \citep{Ishibashi2003}, 
so that the \centconst\  component might be produced by the collision of the 
polar wind from \etacar\  with the Little Homunculus.
The velocity of the polar wind from \etacar\  is believed to be $v \sim$1000~\UNITVEL\ \citep{Smith2003}, 
which is sufficient to heat gas to \KT\  $\sim$1~keV.
If all the kinetic energy of this outflow is converted to thermal energy, the required mass loss rate
would have to be only $\sim$3$\times$10$^{-8}$ \UNITSOLARMASS\  yr$^{-1}$ 
in order to match the observed emission measure ($\sim$7.0$\times$10$^{56}$~\UNITEI),
which is only $3\times10^{-4}-3\times 10^{-5}$ of the mass loss rate of \etacar\ 
\citep[$\sim$10$^{-3}-$10$^{-4}$ \UNITSOLARMASS\  yr$^{-1}$,][]{Hillier2001,Pittard2002}.
The gas density inside the shell structure of the Little Homunculus should be less than
the density of the shell, which is 10$^{6}<n_{\rm H}< $10$^{7}$~\UNITPPCC\  \citep{Ishibashi2003}.
Assuming that radiative cooling dominates, the cooling timescale of the plasma is 
$\tau\sim$3$nkT/n^{2}\Lambda(T)$, where $\Lambda(T)$ is the emissivity.
With a plasma density of  $n<$10$^{7}$~\UNITPPCC,
and plasma temperature \KT\  $\sim$1~keV, and $\Lambda$(1keV) $\sim$ 3.1$\times$10$^{-23}$
ergs s$^{-1}$ cm$^{3}$,
then $\tau >$0.5 years, consistent with our result.

A third alternative is that the \centconst\  might be associated with the outflowing material 
from the WWC that collides with cold circumstellar gas and is reheated.
The required mass loss rate streaming out through the WWC is given by
\Mdot\  = 2\LX/$\epsilon V^{2} \sim$10$^{-8}$/($\epsilon \times (V/2000 $\UNITVEL$)^{2}$)~\UNITSOLARMASS\  yr$^{-1}$,
where $\epsilon$ is the efficiency to convert the kinetic energy to X-ray emission,
and $V$ is the velocity of the outflow, which should be $\gtrsim$1,000~\UNITVEL\  
from the CCE temperature.
Since the most of the companion's wind goes into the shock cone,
the \Mdot\  of the outflowing material should be $\sim$10$^{-5}$~\UNITSOLARMASS\  yr$^{-1}$
\citep[][]{Pittard2002},
and $\epsilon$ should only be $\sim$0.1\%.

\section{Conclusions \& Summary}

We have observed \etacar\  at key phases around the X-ray minimum in 2003
using the \XMM\  and \CHANDRA\  satellites and studied the variation of the
X-ray spectral shape.
We also compared the behavior of the X-ray emission during the 2003 minimum with 
observations during earlier minima made by \ASCA, \SAX\  and \ROSAT.
The results are summarized as follows.

\begin{enumerate}

\item During the minimum, the hard and medium band emission from the central point source decreased 
to 1$-$5\% of the flux outside the minimum, but the emission did not disappear.
The X-ray flux varied in an apparently stepwise way near the middle of the minimum, suggesting that the X-ray
minimum has two states.
The light curve within any single observation did not show significant variability on timescales of a few ksec, supporting that the hard X-ray emission is produced in a WWC.

\item 
The spectrum of the variable emission component changed markedly throughout the X-ray minimum.
The medium band emission decreased gradually relative to the hard emission.
Assuming an absorbed 1T model for the entire spectrum,
the absorption to the central source increased from 5$\times$10$^{22}$~\UNITNH\  to 4$\times$10$^{23}$~\UNITNH.
The hard band flux went up prior to the X-ray minimum, and decreased abruptly during the 2003 minimum, 
without changing the spectral slope significantly,
suggesting that the plasma electron temperature is constant around 3~keV.

\item The line flux ratios between the helium-like ions and hydrogen-like ions of Si, S, and Fe
seen in the spectra before the 2003 minimum indicate a similar type of multi-temperature plasma as
seen near apastron by \citet{Corcoran2001a}.

\item The values of \NH\  derived from the spectra above 5~keV (\NHHOT) are larger than 
those values of \NH\  
derived from the entire spectra (\NHONET) or
from the spectra below 5~keV (\NHCOOL), 
but show less variation than \NHONET\  or \NHCOOL.
This suggests that the hotter plasma has stronger absorption, i.e. emission from \etacar\  
is multi-temperature, and each emission component is associated with an independent absorption component. 
This means that the hotter and cooler emission originate in spatially distinct regions.
The hardest emission, which is believed to arise from material near the stagnation point 
of the WWC region,
perhaps shows the effects of a build-up of material as the colliding wind shock plows through the primary wind.

\item The \ion{Fe}{25} line profile showed a small blue excess, and a strong red excess 
before the 2003 minimum.
The blue excess may be due to Doppler shifting of the line centroid or line broadening,
and if so, the observed velocity is $\sim$4000~\UNITVEL.
The red excess may indicate that some of the plasma is no longer in 
ionization equilibrium.
The line showed a stronger red excess during the 2003 minimum than outside the minimum, 
which we interpret as decrease in the cooling time scale of the plasma near periastron.
Interestingly, we note that an outburst spectrum of the dwarf nova SS Cyg taken with the \SUZAKU\  satellite
showed a similar red excess on the Fe K line profile accompanied with a decrease in the hard band X-ray flux \citep{Ishida2007}.

\item We find evidence for a hard tail at $E>9$ keV, at about an order of magnitude larger than 
the background level, in spectra outside the X-ray minimum.
This either indicates the presence of very hot plasma in the system or non-thermal emission 
perhaps produced by first-order Fermi acceleration of particles which then Compton-upscatter UV 
seed photons from the stellar photospheres and/or X-ray photons from the shock.

\item The values of \NH\  and \KT\  of the X-ray spectrum at corresponding phases in previous cycles
are very similar to those during the 2003 cycle, except during the recovery phase.
In particular, the \ASCA\  spectrum during the 1997/98 minimum (ASCA$_{\rm 971224}$)
is almost identical to the spectrum from CXOXMM$_{\rm 0307}$.
The difference in \NH\  during the recovery, as measured by \CHANDRA\  and \SAX\ 
could  be produced by an increase in the density of the wind from \etacar.

\item The EWs of the Fe fluorescence line at 6.4~keV are always below the values expected
from \NHHOT\  assuming a spherically symmetric absorber.
This is probably because the absorber is not spherically symmetric and is consistent
with the lack of scattering matter in the lower-density wind of the companion if the full opening angle of the shock cone is $\sim120$\DEGREE,
consistent with the value derived by \citet{Pittard2002}.

\item The X-ray minimum is caused by the apparent decrease of the plasma \EM\ 
An eclipse model seems to explain the observed constancy of the plasma electron temperature, while
the small EWs of 
the Fe fluorescence line and the deformation of the \ion{Fe}{25} He-like triplet might indicate a real decline in the amount of hot gas 
in the colliding wind shock.  It might also indicate that nearly all of the shock is hidden by
an optically thick but slightly porous stellar wind.

\item We discovered a constant emission (\centconst) component very near the central X-ray source.
The  \centconst\  component can be seen between 1$-$3 keV during the X-ray minimum; at other times the emission from the colliding wind source overwhelms it.
This component showed no variation on either short timescales within any observation or
between observations for about two months during the 2003 minimum,
and was apparently the same during the 1997/98 and 1992 minima. 
The \centconst\  spectrum exhibited emission lines of \ion{Si}{13} and \ion{S}{14},
and can be fit
by an absorbed 1T model with \NH\  $\sim$5$\times$10$^{22}$~\UNITNH, \KT\  $\sim$1.1~keV
and absorption corrected \LX\  $\sim$10$^{34}$~\UNITLUMI.
\CHANDRA\  data suggest that the emission is point-like, restricting the size of the source 
$\lesssim$2000AU at 2.3~kpc.
This emission might be produced by collisionally-heated shocks from the fast winds from \etacar\  
or perhaps the fast moving outflow from the wind-wind collision with the ambient gas, 
or might be associated with shocked gas which is intrinsic to the wind of \etacar.

\item X-ray emission within 3\ARCMIN\  from \etacar\  can be reproduced by 2 emission components
with interstellar absorption: cool plasma with \KT\  $\sim$0.25~keV (perhaps coming from the 
diffuse emission and HDE~303308), and hot plasma with  \KT\  $\sim$2.6~keV or power-law $\Gamma \sim2.5$,
perhaps coming from a cluster of young stars. Correcting for these emission components is important in understanding the X-ray spectrum of \etacar\  derived from lower spatial resolution X-ray observatories like \ASCA\  and \SUZAKU.

\item Emission from the outer ejecta did not vary significantly on timescales of several years.
The slope of the highest energy emission associated with the spectrum of the outer ejecta can be reproduced with \KT\  $\sim$0.6~keV, 
which is consistent with the highest temperature derived from analysis of the emission line spectrum \citep{Leutenegger2003}.

\end{enumerate}

Currently, there are few reasons to suspect that the variable X-ray emission is not 
driven by shocks produced by the WWC from binary stars.
For the first time, the spatially-resolved observations of \etacar\  obtained by \XMM\  and 
\CHANDRA\  has revealed the details of the plasma and absorption components 
which are directly associated with the WWC plasma.
While the earlier spectral analysis of \citet{Pittard2002}
did not account for the different emission components identified in this
work, their fit was largely determined by emission above 2 keV which
we have shown is dominated by emission from the WWC at the
phase of their analysis.

Still the biggest mystery is the cause of the X-ray minimum, which 
is difficult to fully explain by occultation by a smooth wind from the primary star.
If the X-ray activity really switches off during the minimum, a fading of the X-ray emission
can be imaged in principle using the scattered X-ray emission from the Homunculus Nebula.
The  intensity of the Fe K-shell fluorescence  line just after the onset of the 2003 minimum should
change differently between the eclipsing model and the fading model, 
and perhaps detailed monitoring of this line by spatially resolved X-ray spectral observations could decide the issue.
On the other hand, deformation of the \ion{Fe}{25} line profiles is unusual in the context of 
stellar systems, though similar deformation has been noted in iron lines arising from regions of strong gravity near black holes.
This feature requires further study with high resolution spectrometers such 
as an X-ray micro-calorimeter to produce sufficient spectral resolution at 7~keV.

\acknowledgments

We are grateful to A. Pollock, K. Koyama, K. Misaki and K. Mukai for useful comments.
We greatly appreciate the \XMM\  and \CHANDRA\  review committee's positive response 
for the TOO observation proposal.
This work is performed while K.H. held awards by
National Research Council Research Associateship Award at NASA/GSFC,
and supported by \CHANDRA\  US grant (No. GO3-4008A). This research has made use of NASA's Astrophysics Data System.
This research has made use of data obtained from the High Energy Astrophysics Science Archive
Research Center (HEASARC), provided by NASA's Goddard Space Flight Center.

Facilities: \facility{XMM-Newton(EPIC)}, \facility{CXO(ACIS-S, HETG)}, \facility{ASCA(GIS)}, \facility{ROSAT(PSPC-B)}

\bibliographystyle{apj}
\bibliography{inst,sci_AI,sci_JZ,scibook}

\end{document}

%% file: tab1.tex
\begin{deluxetable}{lclcccc}
\tablecolumns{7}
\tablewidth{0pc}
\tabletypesize{\scriptsize}
\tablecaption{Logs of the {\it XMM-Newton} and {\it Chandra} Observations\label{tbl:obslogs}}
\tablehead{
\colhead{Abbr.}&
\colhead{Seq. ID}&
\multicolumn{2}{c}{Observation Start}&
\colhead{Exposure}&
\colhead{Observation Mode}&
\colhead{Filter}\\ \cline{3-4}
&&\colhead{Date}&\colhead{$\phi_{orbit}$}& \colhead{(ksec)}
}
\startdata
\multicolumn{7}{l}{Before the minimum}\\
XMM$_{\rm 000726}$\tablenotemark{a}&112580601&2000 Jul. 26, 06:01 (M2) &0.470 &\nodata/\nodata/29.6& PFW/PFW/PSW&thick\\
XMM$_{\rm 000728}$\tablenotemark{a}&112580701&2000 Jul. 28, 00:51 (M2) &0.471 &\nodata/\nodata/7.8 & PFW/PFW/PSW&thick\\
XMM$_{\rm 030125}$\tablenotemark{b}&145740101&2003 Jan. 25, 13:03 (PN)&0.922 &4.8/\nodata/\nodata&PSW/PFW/PFW&thick\\
XMM$_{\rm 030127A}$\tablenotemark{b}&145740201&2003 Jan. 27, 01:08 (PN)&0.922 &4.8/\nodata/\nodata&PSW/PFW/PFW &thick\\
XMM$_{\rm 030127B}$\tablenotemark{b}&145740301&2003 Jan. 27, 20:42 (PN)&0.923 &4.7/\nodata/\nodata&PSW/PFW/PFW &thick\\
XMM$_{\rm 030129}$\tablenotemark{b}&145740401&2003 Jan. 29, 01:45 (PN)&0.923 &5.8/\nodata/\nodata&PSW/PFW/PFW &thick\\
XMM$_{\rm 030130}$\tablenotemark{b}&145740501&2003 Jan. 30, 00:00 (PN)&0.924 &4.8/\nodata/\nodata&PSW/PFW/PFW &thick\\
XMM$_{\rm 030608}$&160160101&2003 Jun. 08, 13:31 (M1) &0.988 &22.1/27.6/\nodata&PSW/PSW/PFW &thick\\
XMM$_{\rm 030613}$&160160901&2003 Jun. 13, 23:52 (M1) &0.990 &21.9/30.4/\nodata&PSW/PSW/PFW &thick\\
\multicolumn{7}{l}{During the minimum}\\
CXO$_{\rm 030720}$\tablenotemark{c}&200216&2003 Jul. 20, 01:47 &1.008&90.3&ACIS-S w/HETG&\nodata\\
XMM$_{\rm 030722}$\tablenotemark{c}&145780101&2003 Jul. 22, 01:51 (M2) &1.009 &5.8/8.2/8.4&PSW/PSW/PFW &thick\\
XMM$_{\rm 030802}$&160560101&2003 Aug. 02, 21:01 (M2) &1.015 &12.2/17.1/17.5&PSW/PSW/PFW &thick\\
XMM$_{\rm 030809}$&160560201&2003 Aug. 09, 01:44 (M2) &1.018 &8.7/12.3/12.6&PSW/PSW/PFW &thick\\
XMM$_{\rm 030818}$&160560301&2003 Aug. 18, 15:23 (M2) &1.023 &13.0/18.2/18.6&PSW/PSW/PFW &thick\\
CXO$_{\rm 030828}$&200237&2003 Aug. 28, 17:38 &1.028&18.8&ACIS-S&\nodata\\
\multicolumn{7}{l}{After the minimum}\\
CXO$_{\rm 030926}$&200217&2003 Sep. 26, 22:47 &1.042&70.1&ACIS-S w/HETG&\nodata\\
\enddata
\tablecomments{
Abbr.: abbreviation adopted for each observation.
Seq. ID: sequence identification number of each observation.
Observation Start Date: start time of the detector in the parentheses.
$\phi_{orbit}$ = (JD[observation start] $-$ 2450799.792)/2024 \citep{Corcoran2005}.
Exposure: exposure time excluding the detector deadtime;  for {\it XMM-Newton} observations the exposure times are given for the PN, MOS1 and MOS2 detectors, respectively;
The data sets without exposure time were not used due to the severe event pile-up.
Observation Mode: PSW: prime small window, PFW: prime full window. For the {\it XMM-Newton} observations the observations modes are given for the PN, MOS1 and MOS2 detectors, respectively.
Filter: optical blocking filter selected for the {\it XMM-Newton} detectors.
}
\tablenotetext{a}{Data from XMM$_{\rm 000726}$ and XMM$_{\rm 000728}$ are combined and denoted as XMM$_{\rm 0007}$.}
\tablenotetext{b}{Data from XMM$_{\rm 030125}$, XMM$_{\rm 030127A}$, XMM$_{\rm 030127B}$, XMM$_{\rm 030129}$ and XMM$_{\rm 030130}$ are combined and denoted as XMM$_{\rm 0301}$.}
\tablenotetext{c}{Spectra from CXO$_{\rm 030720}$ and XMM$_{\rm 030722}$ are simultaneously fit and those results are denoted as CXOXMM$_{\rm 0307}$.}
\end{deluxetable}

%% file: tab2.tex
\begin{deluxetable}{lr@{~}c@{~}lr@{~}c@{~}lr@{~}c@{~}lc@{~}c@{~}cc}
\tablecolumns{12}
\tablewidth{0pc}
\tabletypesize{\scriptsize}
\tablecaption{Time Variability\label{tbl:cntslc}}
\tablehead{
\colhead{Observation}&\multicolumn{9}{c}{Net Count Rate}&\multicolumn{4}{c}{Constant Model Fit}\\
&\multicolumn{3}{c}{Soft[0.3$-$1 keV]}&\multicolumn{3}{c}{Med[1$-$4 keV]}&\multicolumn{3}{c}{Hard[4$-$10 keV]}&\multicolumn{4}{c}{$\chi^{2}$/d.o.f.}\\
&\colhead{PN}&\colhead{MOS1}&\colhead{MOS2}&\colhead{PN}&\colhead{MOS1}&\colhead{MOS2}&\colhead{PN}&\colhead{MOS1}&\colhead{MOS2}&\colhead{Soft}&\colhead{Med}&\colhead{Hard}&\colhead{d.o.f}\\
&\multicolumn{3}{c}{({\rm counts~s$^{-1}$})}&\multicolumn{3}{c}{({\rm counts~s$^{-1}$})}&\multicolumn{3}{c}{({\rm counts~s$^{-1}$})}&\multicolumn{4}{c}{}}
\startdata
XMM$_{\rm 0007}$&\nodata&\nodata&0.255\tablenotemark{a}&\nodata&\nodata&1.25\tablenotemark{a}&\nodata&\nodata&0.91\tablenotemark{a}&0.97&1.12&0.82&76\\
XMM$_{\rm 0301}$&1.15&\nodata&\nodata&7.72&\nodata&\nodata&7.18&\nodata&\nodata&0.98&1.14&1.05&75\\
XMM$_{\rm 030608}$&1.10&0.247\tablenotemark{a}&\nodata&4.58&1.70\tablenotemark{a}&\nodata&6.77&1.89\tablenotemark{a}&\nodata&0.99&1.45&0.95&55\\
XMM$_{\rm 030613}$&1.09&0.242\tablenotemark{a}&\nodata&4.99&1.84\tablenotemark{a}&\nodata&10.49&2.80\tablenotemark{a}&\nodata&1.02&1.84&1.29&62\\
XMM$_{\rm 030722}$&1.03&0.223\tablenotemark{a}&0.285&0.456&0.170\tablenotemark{a}&0.188&0.183&0.045\tablenotemark{a}&0.049&0.81&0.85&1.31&16\\
XMM$_{\rm 030802}$&1.07&0.222\tablenotemark{a}&0.288&0.464&0.161\tablenotemark{a}&0.190&0.517&0.137\tablenotemark{a}&0.124&0.97&1.21&0.75&34\\
XMM$_{\rm 030809}$&1.06&0.226\tablenotemark{a}&0.280&0.476&0.178\tablenotemark{a}&0.188&0.554&0.159\tablenotemark{a}&0.151&1.13&0.79&1.09&24\\
XMM$_{\rm 030818}$&1.05&0.221\tablenotemark{a}&0.288&0.514&0.169\tablenotemark{a}&0.202&0.648&0.179\tablenotemark{a}&0.174&1.34&0.96&0.92&37\\
\enddata
\tablenotetext{a}{Source regions are slightly smaller for these Prime Small Window observations.}
\end{deluxetable}

%% file: tab3.tex
\begin{deluxetable}{ccc}
\tablecolumns{3}
\tablewidth{0pc}
\tabletypesize{\scriptsize}
\tablecaption{Best Fit Parameters of the CCE Component \label{tbl:centconst}}
\tablehead{
\colhead{Parameters}&&\colhead{1T fit}
}
\startdata
{\it kT}&(keV)&1.05 (0.96--1.13)\\
{\it Z}(Si)&(solar)&0.22 (0.14--0.29)\\
{\it Z}(S)&(solar)&0.47 (0.39--0.55)\\
{\it E.M.}&(10$^{56}${\rm cm$^{-3}$})&7.0 (6.0--8.2)\\
{\it N$_{\rm H}$}&(10$^{22}${\rm cm$^{-2}$})&4.96 (4.77-5.16)\\
{\it L$_{\rm X}$}\tablenotemark{a}&(10$^{34}${\rm ergs~s$^{-1}$})&1.3\\
$\chi^{2}$/d.o.f&&1.51\\
d.o.f&&1205\\
\enddata
\tablecomments{Elemental abundances of the hard component were found
to be non-solar, but do not vary with observation.}
\tablenotetext{a}{Absorption corrected X-ray luminosity between 0.5--10~keV,
assuming the distance of 2.3~kpc.}
\end{deluxetable}

%% file: tab4.tex
\begin{deluxetable}{lccccccc}
\rotate
\tablecolumns{9}
\tablewidth{0pc}
\tabletypesize{\scriptsize}
\tablecaption{The best-fit models for the whole spectra\label{tbl:specallfit}}
\tablehead{
\colhead{Observation}&\colhead{$\phi_{orbit}$}&\colhead{{\it kT}}&\colhead{$Z$}&\colhead{Flux$_{6.4~keV}$}&\colhead{{\it N$_{\rm H}$}}&\colhead{Observed Flux}&\colhead{reduced $\chi^{2}$ (d.o.f.)}\\
&&\colhead{(keV)}&\colhead{(solar)}&\colhead{(10$^{-4}${\rm counts~cm$^{-2}$~s$^{-1}$})}&\colhead{(10$^{22}$~{\rm cm$^{-2}$})}&\colhead{(10$^{-11}${\rm ergs~cm$^{-2}$~s$^{-1}$})}
}
\startdata
XMM$_{\rm 0007}$&0.470--0.471& 4.6~(4.5--4.8)&0.67~(0.63--0.70)&0.91~(0.75--1.04)&5.1~(5.0--5.2)& 6.5&2.85 (315)\\
XMM$_{\rm 0301}$&0.922--0.924&  4.3~(4.2--4.4)&0.64~(0.62--0.65)&2.4~(2.3--2.6)   &4.8~(4.7--4.9)&15.0&10.29 (249)\\
XMM$_{\rm 030608}$&0.988&   4.5~(4.4--4.6)&0.83~(0.81--0.85)&4.1~(3.9--4.2)   &7.4~(7.3--7.5)&12.8&12.77 (417)\\
XMM$_{\rm 030613}$&0.990&   5.4~(5.4--5.5)&0.95~(0.94--0.97)&6.9~(6.7--7.1)   &9.1~(9.0--9.2)&19.3&17.49 (548)\\
CXOXMM$_{\rm 0307}$&1.008--1.009&3.1~(2.8--3.6)&0.44~(0.37--0.53)&0.16~(0.13--0.20)&28~(25--32)&0.19&1.24 (193)\\
XMM$_{\rm 030802}$&1.015&   3.6~(3.3--3.9)&0.64~(0.56--0.71)&0.65~(0.58--0.78)\tablenotemark{b}&42~(39--45)&0.75&2.44 (119)\\
XMM$_{\rm 030809}$&1.018&  2.9~(2.8--3.1)&0.64~(0.58--0.71)&0.55~(0.45--0.65)\tablenotemark{b}&36~(34--38)&0.82&1.19 (94)\\
XMM$_{\rm 030818}$&1.023&  2.9~(2.8--3.1)&0.58~(0.54--0.63)&0.37~(0.26--0.42)\tablenotemark{b}&29~(27--30)&0.96&1.30 (161)\\
CXO$_{\rm 030828}$\tablenotemark{a}&1.028&   4.6~(3.4--5.4)&0.48~(0.37--0.60)&0.31~(0.3--0.43)&18~(17--21)&0.73&1.54 (137)\\
CXO$_{\rm 030926}$&1.042&   2.9~(2.3--3.3)&0.40~(0.36--0.45)&3.6~(3.0--4.3)   &36~(33--40)&6.6&1.40 (242)\\
\enddata
\tablecomments{These fits refer toÊ
the spectrum between 2--10~keV after subtraction of the CCE and the Homunculus emission.
Parentheses show the range of values for which $\Delta\chi^{2}=2.7$ from 
the best-fit value.}
\tablenotetext{a}{About 4\% of the photon events are piled-up, causing the spectrum to appear slightly harder.}
\tablenotetext{b}{These values are upper-limits since the line shapes are not clearly seen in the spectra.}
\end{deluxetable}

%% file: tab5.tex
\begin{deluxetable}{lcccccc}
\tablecolumns{7}
\tablewidth{0pc}
\tabletypesize{\scriptsize}
\tablecaption{The best-fit models for the spectra above 5~keV\label{tbl:spechardfit}}
\tablehead{
\colhead{Observation}&\colhead{$\phi_{orbit}$}&\colhead{$Z_{\rm Fe}$}&\colhead{log {\it E.M.}}&EW(Fe)&\colhead{{\it N$_{\rm H}$}}&\colhead{reduced $\chi^{2}$ (d.o.f.)}\\
&&\colhead{(solar)}&\colhead{({\rm cm$^{-3}$})}&\colhead{(eV)}&\colhead{(10$^{22}$~{\rm cm$^{-2}$})}
}
\startdata
XMM$_{\rm 0007}$&0.470--0.471&0.48~(0.45--0.52)&58.1~(58.1$-$58.2)&87~(62--99)&17~(14--20)&1.43~(86)\\
XMM$_{\rm 0301}$&0.922--0.924&0.48~(0.47--0.50)&58.5~(58.5$-$58.5)&99~(85--104)&17~(16--19)&4.07~(82)\\
XMM$_{\rm 030608}$&0.988&0.55~(0.54--0.57)&58.5~(58.5$-$58.5)&164~(158--172)&21~(20--22)&4.21~(154)\\
XMM$_{\rm 030613}$&0.990&0.44~(0.43--0.45)&58.9~(58.9$-$58.9)&119~(114--123)&36~(35--37)&3.66~(245)\\
CXOXMM$_{\rm 0307}$&1.008--1.009&0.38~(0.30--0.47)&57.0~(56.9$-$57.0)&$\sim$233~(183--325)\tablenotemark{a}&40~(32--51)&1.35~(90)\\
XMM$_{\rm 030802}$&1.015&0.50~(0.46--0.55)&57.7~(57.6$-$57.7)&$\lesssim$213\tablenotemark{a}&53~(48--58)&1.91~(71)\\
XMM$_{\rm 030809}$&1.018&0.70~(0.63--0.77)&57.5~(57.4$-$57.5)&$\lesssim$257\tablenotemark{a}&32~(26--37)&1.59~(54)\\
XMM$_{\rm 030818}$&1.023&0.62~(0.57--0.67)&57.5~(57.5$-$57.5)&$\lesssim$150\tablenotemark{a}&27~(23--30)&1.65~(84)\\
CXO$_{\rm 030828}$\tablenotemark{b}&1.028&0.33~(0.24--0.42)&57.5~(57.4$-$57.6)&$\sim$149~(96--214)\tablenotemark{a}&37~(28--47)&1.81~(54)\\
CXO$_{\rm 030926}$&1.042&0.33~(0.29--0.37)&58.7~(58.6$-$58.8)&110~(90--136)&56~(50--62)&1.54~(130)\\
\enddata
\tablecomments{
These fits refer to the spectrum after subtraction of the CCE and the Homunculus emission.
The plasma temperature and the Ni abundance were fixed at 3.3~keV and 0.8~solar, respectively. 
}
\tablenotetext{a}{These values are regarded as upper-limits since
the spectra did not clearly show the iron fluorescence line peak.}
\tablenotetext{b}{About 4\% of the photon events pile-up, causing the spectrum slightly hard.}
\end{deluxetable}

%% file: tab6.tex
\begin{deluxetable}{ccccccc}
\tablecolumns{7}
\tablewidth{0pc}
\tabletypesize{\scriptsize}
\tablecaption{Additional soft component for the spectrum below 5~keV\label{tbl:spec2tfit}}
\tablehead{
\colhead{Observation}&\colhead{$\phi_{orbit}$}&\colhead{{\it kT}}&\colhead{$Z$}&\colhead{log {\it E.M.}}&\colhead{{\it N$_{\rm H}$}}&\colhead{Reduced $\chi^{2}$ (d.o.f.)}\\
&&\colhead{(keV)}&\colhead{(solar)}&\colhead{({\rm cm$^{-3}$})}&\colhead{(10$^{22}$~{\rm cm$^{-2}$})}
}
\startdata
XMM$_{\rm 0007}$&0.470--0.471&1.1~(1.1$-$1.2)  &0.25~(0.20$-$0.31)&58.3~(58.2$-$58.4)&5.4~(5.2$-$5.5)&2.04~(225)\\
XMM$_{\rm 0301}$&0.922--0.924&1.1~(1.1$-$1.1)  &0.18~(0.16$-$0.19)&58.8~(58.8$-$58.8)&5.4~(5.3$-$5.4)&8.85~(164)\\
XMM$_{\rm 030608}$&0.988&1.1~(1.1$-$1.1)&0.24~(0.22$-$0.26)&58.6~(58.6$-$58.7)&6.9~(6.8$-$7.0)&5.41~(260)\\
XMM$_{\rm 030613}$&0.990&1.6~(1.5$-$1.6)&0.25~(0.22$-$0.28)&58.6~(58.5$-$58.6)&7.7~(7.5$-$7.8)&12.02~(300)\\
CXO$_{\rm 030828}$&1.028&0.5~(0.4$-$0.9)&$<$0.58           &59.9~(58.1$-$60.1)&29~(17$-$39)&1.32~(80)\\
CXO$_{\rm 030926}$&1.042&0.9~(0.6$-$1.8)&0.28~(0.05$-$0.75)&59.2~(58.0$-$60.5)&23~(13$-$33)&0.52~(109)\\
\enddata
\tablecomments{These fits refer to the spectrum after subtraction of the CCE and the Homunculus emission.}
\end{deluxetable}

%% file: tab7.tex
\begin{deluxetable}{lccccccc}
\tablecolumns{8}
\tablewidth{0pc}
\tabletypesize{\scriptsize}
\tablecaption{Observations around the Minimum From Previous Cycles
\label{tbl:obslogsprecycle}}
\tablehead{
\colhead{Abbreviation}&\colhead{Satellite}&\colhead{Date}&\colhead{P$_{orbit}$}&\colhead{Exposure}&\colhead{{\it N$_{\rm H}$}}&\colhead{{\it kT}}&\colhead{Referecnce}\\
&&&&(ksec)&(10$^{22}$~{\rm cm$^{-2}$})&(keV)&
}
\startdata
ROSAT$_{\rm 920612}$&{\it ROSAT}&1992 Jun. 12, 22:33&$-$0.998&23.6&\nodata&\nodata&\nodata\\
ASCA$_{\rm 970703}$&{\it ASCA}&1997 Jul. 3, 09:58&$-$0.082&12.6&4.16$\pm$0.23&3.79$\pm$0.31&1\\
ASCA$_{\rm 970719}$&{\it ASCA}&1997 Jul. 19, 15:11&$-$0.074&12.8&4.09$\pm$0.30&4.12$\pm$0.39&1\\
ASCA$_{\rm 971224}$&{\it ASCA}&1997 Dec. 24, 09:29&0.004&58.8&\nodata&\nodata&1\\
SAX$_{\rm 980318}$&{\it BeppoSAX}&1998 Mar. 18, 05:01&0.045&39.0\tablenotemark{a}&15.4$\pm$0.4&4.35$\pm$0.15&2\\
ASCA$_{\rm 980716}$&{\it ASCA}&1998 Jul. 16, 04:40&0.104&13.4&5.00$\pm$0.52&4.89$\pm$0.88&1\\
\enddata
\tablecomments{{\it N$_{\rm H}$} and {\it kT} refer to the hottest components in 
2-temperature fits of the entire spectra; {\it ASCA} results refer to analysis of the GIS2 spectra.
Ref. 1:\citet{Corcoran2000}, 2:\citet{Viotti2002}}
\tablenotetext{a}{Exposure time for the MECS23 detector.}
\end{deluxetable}

%% file: tab8.tex
\begin{deluxetable}{lccccccccc}
\rotate
\tablecolumns{10}
\tablewidth{0pc}
\tabletypesize{\scriptsize}
\tablecaption{The spectral fit of the surrounding region\label{tbl:specsurround}}
\tablehead{
&\multicolumn{2}{c}{Soft Component}&&\multicolumn{3}{c}{Hard Component}\\
\cline{2-3}\cline{5-7}
\colhead{Model}&\colhead{{\it kT}}&\colhead{log {\it E.M.}}&&\colhead{{\it kT}}&\colhead{log {\it E.M.}}&\colhead{$\Gamma$}&\colhead{{\it Z}}&\colhead{{\it N$_{\rm H}$}}&\colhead{reduced $\chi^{2}$ (d.o.f.)}\\
&\colhead{(keV)}&\colhead{({\rm cm$^{-3}$})}&&\colhead{(keV)}&\colhead{({\rm cm$^{-3}$})}&&\colhead{(10$^{-1}$~solar)}&\colhead{(10$^{21}$~{\rm cm$^{-2}$})}
}
\startdata
2T	&0.24 (0.18--0.29)&57.7 (57.2--58.5)&&2.6 (1.9--3.3)&56.2 (56.2--56.4)&\nodata&0.36 (0.19--0.70)\tablenotemark{a}&3.6 (2.6--5.5)&1.28 (92)\\
1T + pl &0.27 (0.23--0.30)&56.5 (56.3--56.8)&&\nodata&\nodata&2.5 (2.4--2.7)&3 (fix)\tablenotemark{b}&2.6 (2.1--3.3)&1.26 (93)\\
\enddata
\tablenotetext{a}{The abundace parameters of both soft and hard components are tied together.}
\tablenotetext{b}{The absorption is insensitive to abundance in the range between 0 and 10~solar.}
\tablecomments{Values in parentheses denote 90\% confidence limits.}
\end{deluxetable}